\documentclass[]{aa}
\usepackage[T1]{fontenc}
\usepackage[utf8]{inputenc}

\usepackage{graphicx}
\usepackage{natbib}
\usepackage{ulem}
\usepackage{textcomp}
\usepackage{gensymb}
\usepackage{longtable}
\usepackage{threeparttable}
\usepackage{multicol}
\usepackage{multirow}
\usepackage{textgreek}
\usepackage{float}
\usepackage{todonotes}
\usepackage[Symbol]{upgreek}
\usepackage{amsmath}
\usepackage{etoolbox}
\usepackage{txfonts}
\usepackage{url}
\usepackage{xcolor}

\usepackage[most]{tcolorbox}
\usepackage{hyperref}
\usepackage{xcolor}
\hypersetup{
  colorlinks   = true, 
  urlcolor     = blue, 
  linkcolor    = blue, 
  citecolor   = blue 
}
\usepackage[all]{hypcap} 
\def \gtw{\>\hbox{\lower.25em\hbox{$\buildrel >\over\sim$}}\>}
\def \ltw{\>\hbox{\lower.25em\hbox{$\buildrel <\over\sim$}}\>}

\begin{document} 
	\title{Understanding the radio relic emission in the galaxy cluster MACS\,J0717.5+3745: spectral analysis}
	\titlerunning{MACS J0717.5+3745: spectral index and curvature analysis}
	\authorrunning{Rajpurohit et al.}

\author{K. Rajpurohit\inst{1,2,3}, D. Wittor\inst{4,1}, R. J. van Weeren\inst{5}, F. Vazza\inst{1,2,4}, M. Hoeft\inst{3}, L. Rudnick\inst{6}, N. Locatelli\inst{1,2}, J. Eilek\inst{7,8}, W. R. Forman\inst{9}, A. Bonafede\inst{1,2,4}, E. Bonnassieux\inst{1,2},  C. J. Riseley\inst{1,2}, M. Brienza\inst{1,2}, G. Brunetti\inst{2}, M. Br\"uggen\inst{4}, F. Loi\inst{10}, A. S. Rajpurohit\inst{11}, H. J. A. R\"ottgering\inst{5}, A. Botteon\inst{5}, T. E. Clarke\inst{12}, A. Drabent\inst{3}, P. Dom\'{i}nguez-Fern\'{a}ndez\inst{4}, G. Di Gennaro\inst{5}, and F. Gastaldello\inst{13}} 

\institute{Dipartimento di Fisica e Astronomia, Universit\'at di Bologna, via P. Gobetti 93/2, 40129, Bologna, Italy\\
 {\email{kamlesh.rajpurohit@unibo.it}}
\and
INAF-Istituto di Radio Astronomia, Via Gobetti 101, 40129 Bologna, Italy
\and
Th\"uringer Landessternwarte, Sternwarte 5, 07778 Tautenburg, Germany
\and
Hamburger Sternwarte, Universit\"at Hamburg, Gojenbergsweg 112, 21029, Hamburg, Germany
\and
Leiden Observatory, Leiden University, P.O. Box 9513, 2300 RA Leiden, The Netherlands
\and
Minnesota Institute for Astrophysics, University of Minnesota, 116 Church St. S.E., Minneapolis, MN 55455, USA
\and
Department of Physics, New Mexico Tech, Socorro, NM 87801, USA
\and
National Radio Astronomy Observatory, Socorro NM 87801 USA
\and
Harvard-Smithsonian Center for Astrophysics, 60 Garden Street, Cambridge, MA 02138, USA
\and
INAF-Osservatorio Astronomico di Cagliari, Via della Scienza 5, 09047 Selargius (CA), Italy
\and
Astronomy \& Astrophysics Division, Physical Research Laboratory, Ahmedabad 380009, India
\and
U.S. Naval Research Laboratory, 4555 Overlook Avenue SW, Washington, D.C. 20375, USA
\and
INAF-IASF Milano, via A. Corti 12, 20133 Milano, Italy
}
   
\abstract
{Radio relics are diffuse, extended synchrotron sources that originate from shock fronts generated during cluster mergers. The massive merging galaxy cluster MACS\,J0717.5+3745 hosts one of the more complex relics known to date. We present upgraded Giant Metrewave Radio Telescope band\,3 (300-500\,MHz) and band\,4 (550-850\,MHz) observations. These new observations, combined with published VLA  and the new LOFAR HBA data, allow us to carry out a detailed, high spatial resolution spectral analysis of the relic over a broad range of frequencies. The integrated spectrum of the relic closely follows a power-law between 144\,MHz and 5.5\,GHz with a mean spectral slope $\alpha=-1.16\pm0.03$. Despite its complex morphology, the subregions of the relic and the other isolated filaments also follow power-law behaviors, and show similar spectral slopes. Assuming Diffusive Shock Acceleration, we estimate a dominant Mach number of $\sim 3.7$ for the shocks that make up the relic. Comparison with recent numerical simulations suggests that in the case of radio relics, the slopes of the integrated radio spectra are determined by the Mach number of the accelerating shock, with $\alpha$ nearly constant,  namely between $-1.13$ and $-1.17$, for Mach numbers $3.5 - 4.0$. The spectral shapes inferred from spatially resolved regions show curvature, we speculate that the relic is inclined along the line-of-sight. The locus of points in the simulated color-color plots changes significantly with the relic viewing angle. We conclude that projection effects and inhomogeneities in the shock Mach number dominate the observed spectral properties of the relic in this complex system. Based on the new observations we raise the possibility that the relic and a narrow-angle-tailed radio galaxy are two different structures projected along the same line-of-sight.}


\maketitle
  
\section{Introduction}
 \label{sec:intro}

Merging galaxy clusters often show spectacular, large-scale radio emission, referred to as radio relics and radio halos \citep[see][for a recent review]{vanWeeren2019}. The emission is not associated with individual galaxies in the cluster. The radio spectra of such sources are usually steep{\footnote{$S_{\nu}\propto\nu^{\alpha}$, with spectral index $\alpha$}} ($\alpha \leq-1$). 
Both relics and halos are associated with merging clusters \citep[e.g.,][]{Giacintucci2008,Cassano2010,Finoguenov2010,vanWeeren2011}, suggesting that cluster mergers play a major role in their formation. But the exact formation mechanism of radio relics and halos is not  fully understood. It is thought that part of the kinetic energy that is dissipated during cluster mergers is channeled into the acceleration of particles, via  shocks and turbulence, into the amplification of intracluster magnetic fields \citep{Brunetti2014}.

Radio relics are found in the cluster periphery and often show irregular morphologies. They trace shock waves occurring in the intracluster medium (ICM) during cluster merger events \citep{Sarazin2013,Ogrean2013,vanWeeren2016a,Botteon2016a,Botteon2016b,Urdampilleta2018}. Originally, it was believed that cosmic ray electrons (CRe), which form the radio relics via synchrotron emission, originate from a first-order Fermi process, namely diffusive shock acceleration (DSA) \citep{Roettiger1999,Ensslin1998,Hoeft2007} from the thermal pool. It is currently debated if all relics are caused by re-acceleration of fossil electrons \citep{Bonafede2014,Kang2016a,2019arXiv190700966B}. In this scenario, weaker shocks accelerate a population of aged CRe. For the relic in Abell\,3411-3412, \cite{vanWeeren2017b} provided convincing evidence that part of the relic emission is related to a nearby Active Galactic Nucleus (AGN). 

In a number of relics, high-resolution observations provided evidence of filamentary emission in relics \citep{Owen2014,vanWeeren2017b,Rajpurohit2018,Gennaro2018}. The existence of filaments in relics is not entirely unexpected as simple magneto-hydrodynamical simulations of cluster merger shocks produced numerous structures with a non-constant shock Mach number across the shock surface \citep{Vazza2012,Skillman2013,2019arXiv190911329W,Paola2020}. However, the increasing evidence for small-scale ($5-30$\,kpc) radio filaments with complex morphology is more challenging to explain \citep{vanWeeren2017b,Rajpurohit2018}. 

Wideband radio studies of morphology, spectrum, and curvature distributions  provide crucial insights about the particle acceleration in radio relics and their origins. A recent spectral study of the relic in 1RXS\,J0603+4214 (hereafter ``Toothbrush relic") suggests that the spectral properties of radio relics are dominated by inhomogeneities in Mach number, magnetic field strength, and projection effects \citep{Rajpurohit2020a}. If more relics reveal a similar spectral behavior, this is likely to provide a key insight for understanding of radio relics. 

\section{MACS J0717.5+3745}
 \label{target}

The galaxy cluster MACS\,J0717.5+3745, located at $z =0.5458$, is a complex, massive, and rather unusual system. It is one of the most disturbed, luminous, and hottest clusters known to date. The cluster was discovered by \citet{Edge2003} as part of the MAssive Cluster Surveys \cite[MACS;][]{Ebeling2001}. The cluster has been studied extensively across the electromagnetic spectrum.

Optical and X-ray observations of MACS\,J0717.5+3745 revealed that it consists of at least four components belonging to different merging sub-clusters \citep{Limousin2016,vanWeeren2017b}. Recently \cite{Jauzac2018} found that the cluster is dominated by 9 group scale-structures. The X-ray luminosity of  the cluster is $L_{\rm x, \rm 0.1-2.4\,keV}=2.4\times10^{45}\,\rm erg\,s^{-1}$ with an overall ICM temperature of $12.2\pm0.4\rm\,keV$ \citep{Ebeling2007,vanWeeren2017b}. The bar-shaped structure in the south-east of the cluster is extremely hot with a temperature $\geq20\rm\,keV$ \citep{vanWeeren2017b}. However, no evidence has been found of density and/or temperature discontinuities, which would indicate the presence of shocks \citep{vanWeeren2017b}.

In the radio band, the most prominent and noticeable feature is a spectacular ``chair-shaped" relic \citep[see Fig.\,\ref{XrayR}:][]{Bonafede2009a,vanWeeren2009,PandeyPommier2013,vanWeeren2017b} with an embedded Narrow Angle Tail (NAT) galaxy; labeled as NAT in Fig.\,\ref{XrayR}. High-resolution VLA images reveal that the relic has several filaments on scales down to $\sim30\,\rm kpc$. The relic is polarized at 4.9\,GHz and the degree of polarization has been shown to vary significantly across the northern part of the relic \citep{Bonafede2009a}. The southern part of the relic is reported to be apparently associated with the NAT, indicating that this part of the relic is seeded by the re-acceleration of the electrons injected into the ICM by the AGN jet activity \citep{vanWeeren2017b}. MACS\,J0717.5+3745 also hosts a powerful radio halo. In fact, it is one of the most complex and powerful relic-halo systems observed. Our paper focuses on the radio relic emission.

In this work, we describe the results of observations of the galaxy cluster MACS\,J0717.5+3745 with the upgraded Giant Metrewave Radio Telescope (uGMRT). These observations are complemented by the published Karl G. Jansky Very Large Array (VLA) L, S, C-band, and the new LOw-Frequency ARray (LOFAR) observations (Rajpurohit et al. submitted) to perform a high-resolution spectral analysis. The outline of this paper is as follows: In Sect.\,\ref{obs}, we describe the observations and data reduction. The total power and polarization images are presented in Sect.\,\ref{results}. This is followed by a detailed analysis and discussion from Sects.\,\ref{spectral} to \,\ref{simlation_part}. We summarize our main findings in Sect.\,\ref{summary}. 

Throughout this paper we assume a $\Lambda$CDM cosmology with $H_{\rm{ 0}}=70$ km s$^{-1}$\,Mpc$^{-1}$, $\Omega_{\rm{ m}}=0.3$, and $\Omega_{\Lambda}=0.7$. At the cluster's redshift, $1\arcsec$ corresponds to a physical scale of 6.4\,kpc. All output images are in the J2000 coordinate system and are corrected for primary beam attenuation.


\section{Observations and Data Reduction}
\label{obs}

\subsection{uGMRT band 4 and band 3 observations}

The GMRT observations of MACS\,J0717+3745 were carried out with the upgraded GMRT (uGMRT) in band\,4 (proposal code: 36\_006) and band 3 (proposal code: 31\_037). The total bandwidth was 400\,MHz for the band\,3 and 200\,MHz for the band\,4; see Table\,\ref{Tabel:obs} for observational details. The primary calibrators 3C138 and 3C147 were included to correct for the bandpass. The secondary calibrators 3C286 and J0731+438 were used for phase calibration. 

The data were calibrated with $\tt{CASA}$, version 4.7.0. We split the band\,4 data into four sets, namely set1(channels\,$0{-}999$), set2 (channels\,$1000{-}1999$), set3 (channels\,$2000{-}2999$) and set4 (channels\,$3000{-}4096$). The data in set4 were heavily affected by radio frequency interference (RFI), and were thus fully flagged out. Similarly, the band3 data was split into two sets. 
These data sets were processed independently using the steps described hereafter. We first visually inspected data for the presence of RFI which were subsequently removed using ${\tt AOFlagger}$ \citep{Offringa2010}. 

\setlength{\tabcolsep}{5pt}
\begin{table}[!htbp]
\caption{Observational overview: uGMRT observations}
\centering
\begin{threeparttable} 
\begin{tabular}{ l  c  c c}
  \hline  \hline  
Observation date& September 1, 2019  & March 11, 2017  \\
Frequency range&550-950\,MHz&300-500\,MHz\\ 
Correlations &RR, LL, RL, LR &RR, LL\\
Channel width & 49\,kHz & 97\,kHz\\ 
No of channels &4096 &2048\\ 
Integration time& 5\,s & 4\,s  \\
On source time &8\,hrs &6\,hrs  \\
\hline 
\end{tabular}
\end{threeparttable} 
\label{Tabel:obs}   
\end{table}

\begin{figure}[!thbp]
    \centering
    \includegraphics[width=0.49\textwidth]{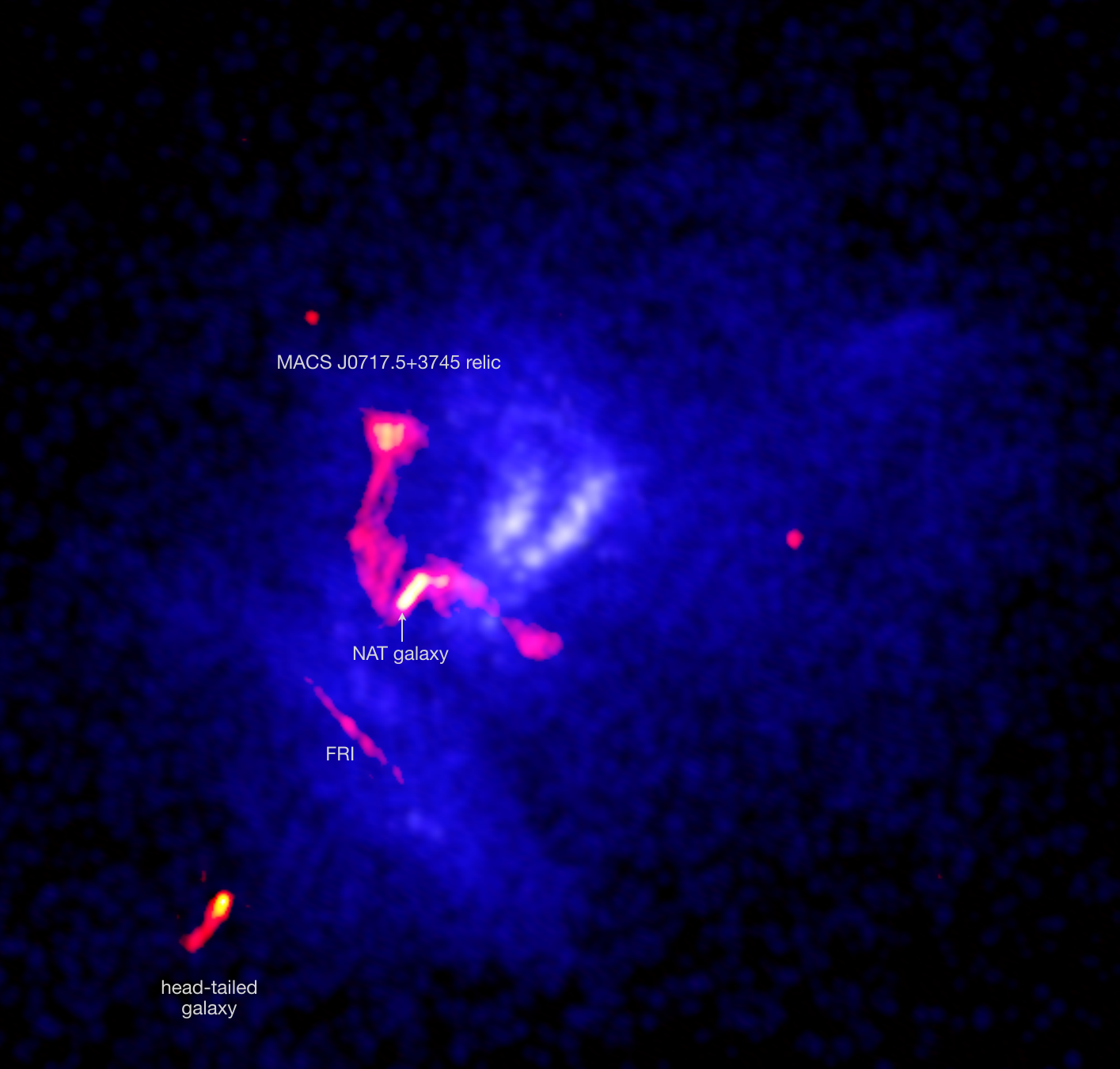}   
    \vspace{-0.1cm}
 \caption{Radio and X-ray overlay of the galaxy cluster MACS\,J0717+3745. The intensity in red shows the radio emission observed with uGMRT band 4 at a central frequency of 700\,MHz. The intensity in blue shows Chandra X-ray emission in the 0.5-2.0 keV band \citep{vanWeeren2017b}. The radio image is obtained using {\tt Uniform} weighting and has a beam size of $3.5\arcsec\times3.5\arcsec$. The noise level is  $\rm \sigma_{rms}=28\,\upmu Jy\,beam^{-1}$. }
      \label{XrayR}
  \end{figure}

\begin{figure*}[!thbp]
    \centering
     \includegraphics[width=0.49\textwidth]{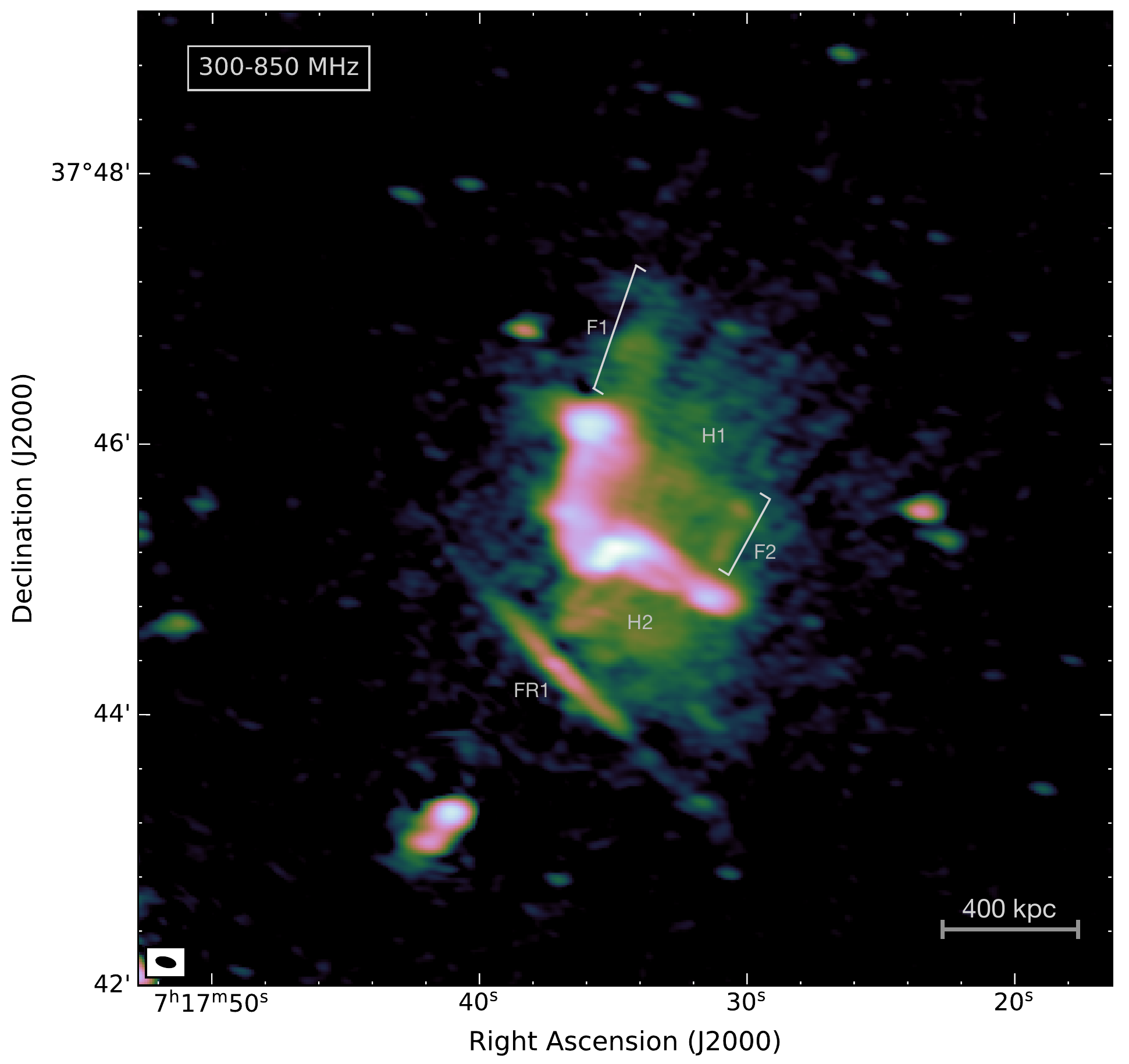}    
    \includegraphics[width=0.49\textwidth]{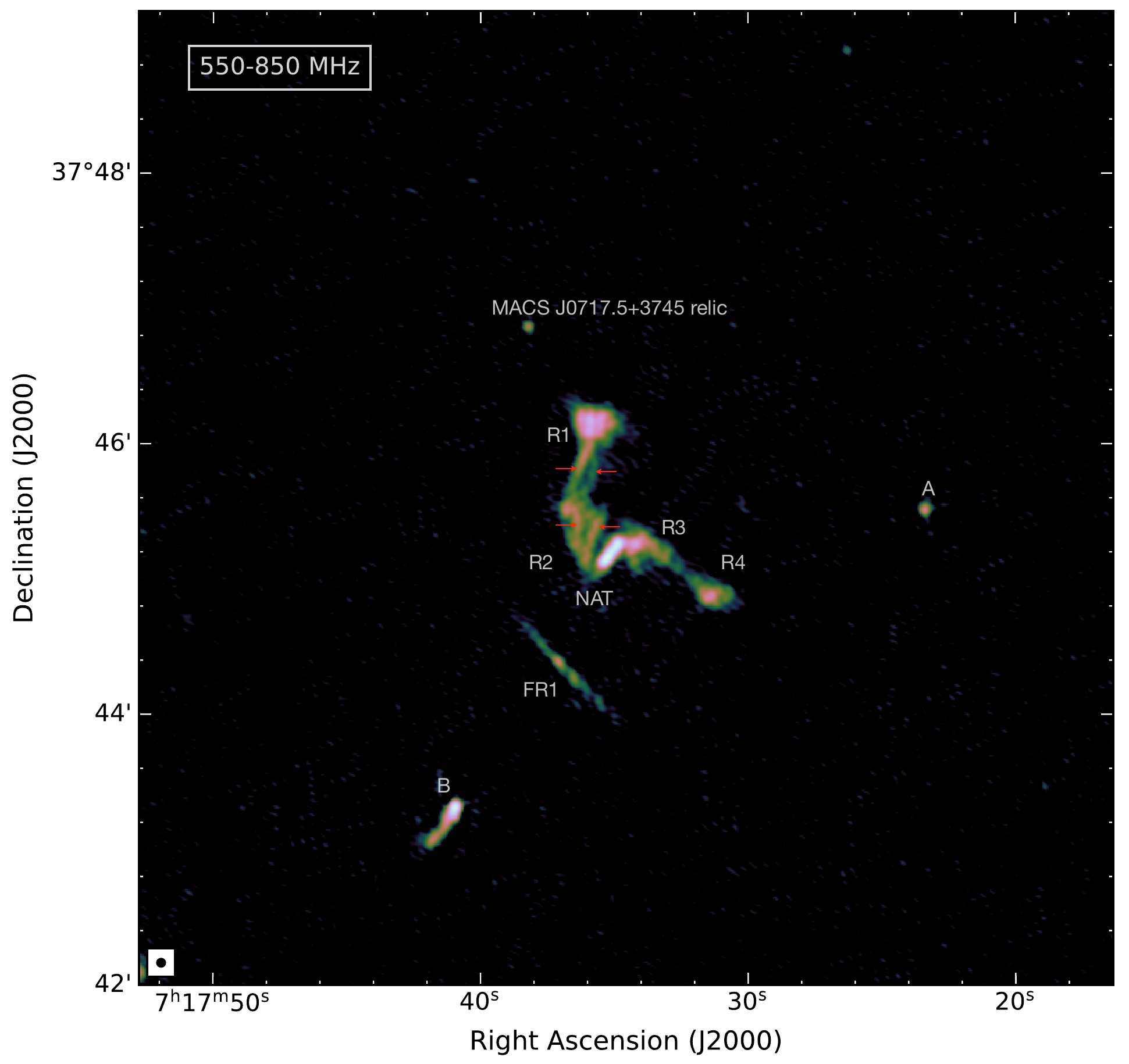}    
    \vspace{-0.1cm}
 \caption{uGMRT  images of the galaxy cluster MACS\,J0717+3745. Images show the "chair-shaped" relic and filamentary structures also reported at high frequency \citep{vanWeeren2017b}. The image demonstrate that the northern part of the relic is composed of multiple filaments (shown with red arrows). \textit{Left}: The combined uGMRT band 3 (300-500\,MHz) and band 4 (550-850 MHz) image, created using {\tt Briggs} weighting and ${\tt robust}=0$. The image has a beam size of $8\arcsec\times4\arcsec$ and noise level of $\rm \sigma_{rms}=13\,\upmu Jy\,beam^{-1}$. \textit {Right}: High resolution uGMRT band\,4 image obtained using {\tt Uniform} weighting and has a beam size of $3.5\arcsec\times3.5\arcsec$. The noise level is  $\rm \sigma_{rms}=28\,\upmu Jy\,beam^{-1}$. The beam shape is indicated in the bottom left corner of each image. }
      \label{fig1}
  \end{figure*}   

\begin{figure*}[!thbp]
    \centering
 \includegraphics[width=0.49\textwidth]{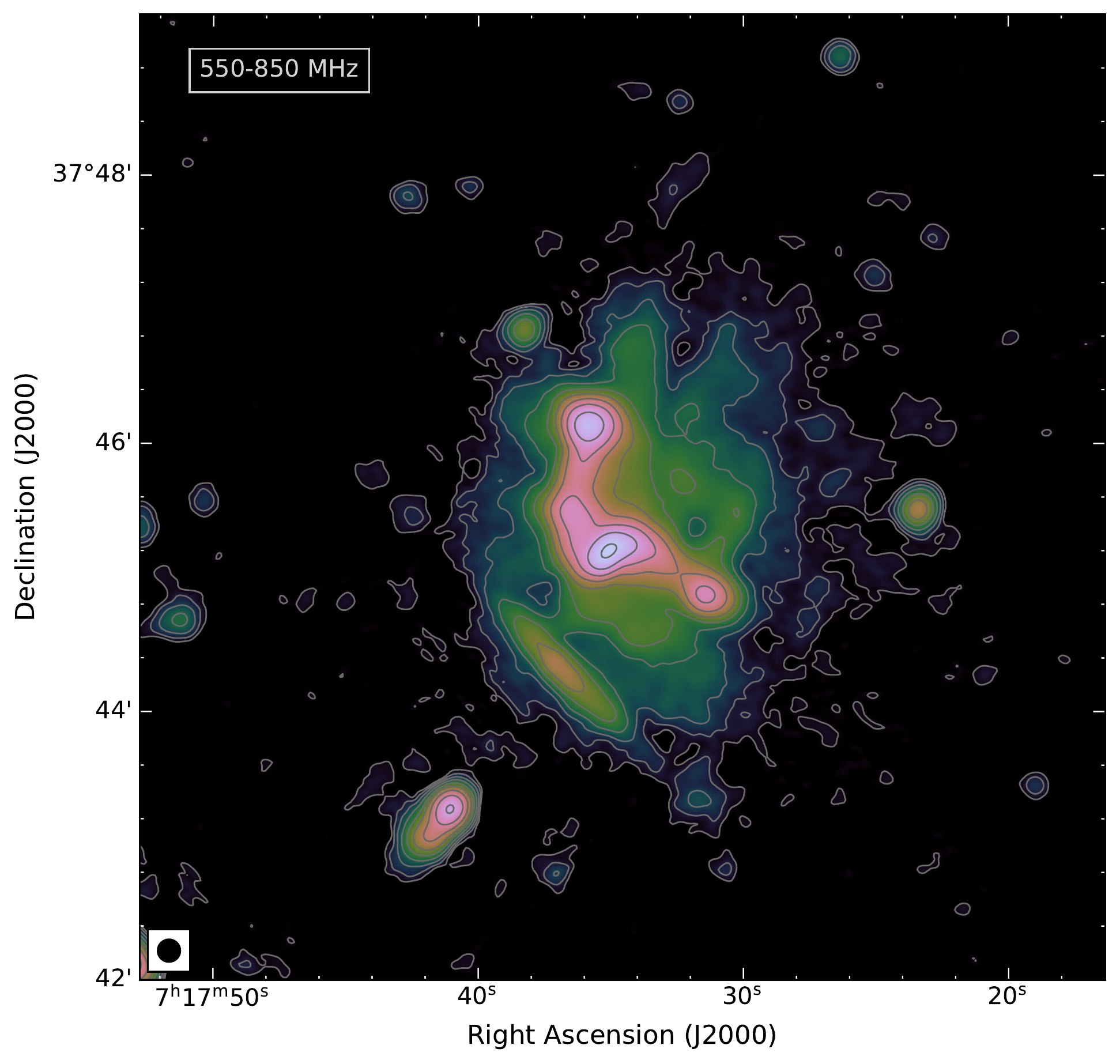}    
  \includegraphics[width=0.49\textwidth]{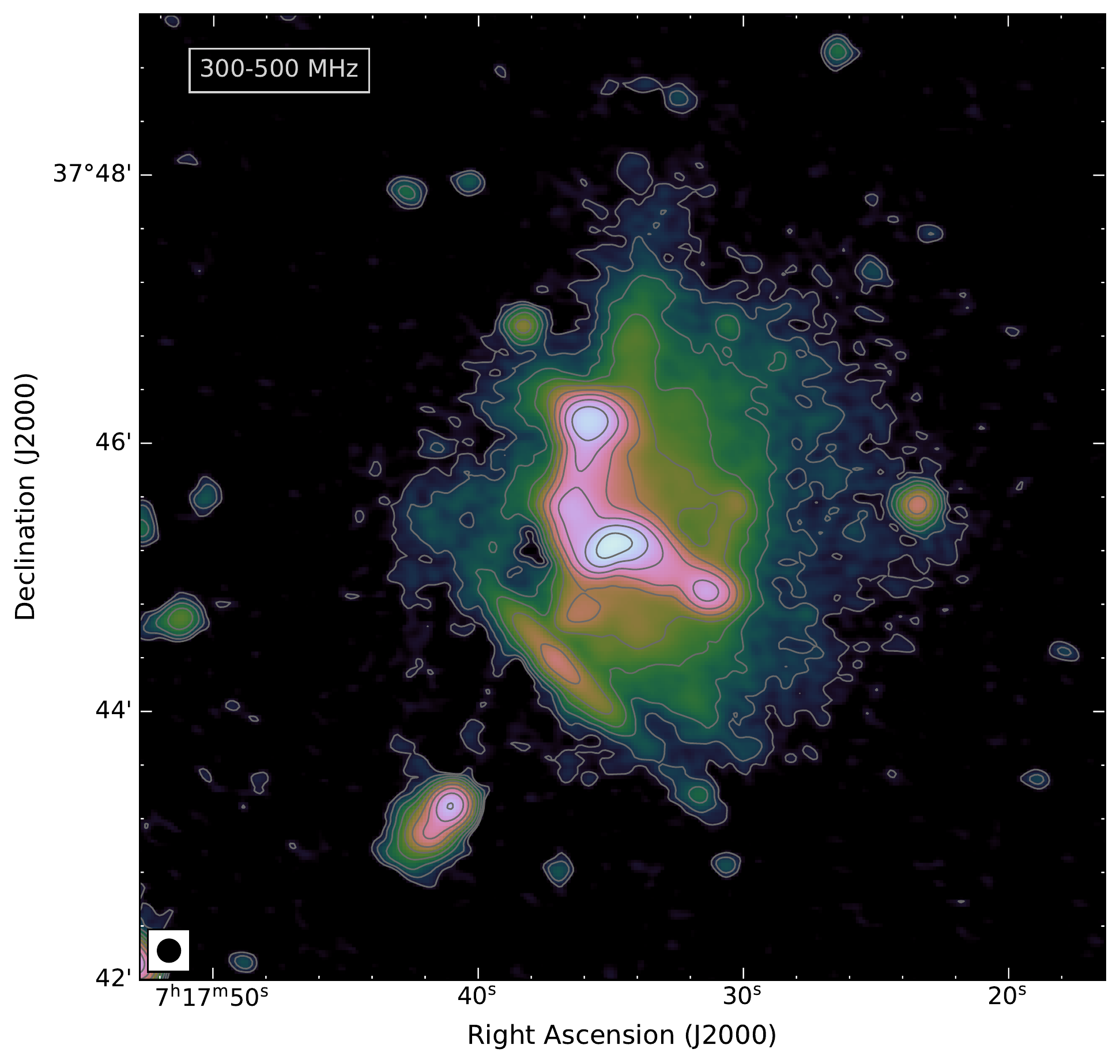}    
     \vspace{-0.1cm}
 \caption{Low resolution uGMRT images of the galaxy cluster MACS\,J0717+3745 showing the large scale halo emission. The halo emission is more extended to the north and to the west than previously observed. \textit{Left}: uGMRT 550-850 MHz image, it has a resolution of $10\arcsec$ and a noise level of $\rm \sigma_{rms, \,700\,MHz}=16\,\upmu Jy\,beam^{-1}$. \textit{Right}: uGMRT 300-500 MHz image, it has a beam size of $10\arcsec$ and a noise level of $\rm \sigma_{rms,\,400\,MHz}=27\,\upmu Jy\,beam^{-1}$. Contour levels are drawn at $[1,2,4,8,\dots]\,\times\,4.0\,\sigma_{{\rm{ rms}}}$. In these images there is no region below $-4.0\,\sigma_{{\rm{ rms}}}$. Images are obtained using ${\tt Briggs}$ weighting with ${\tt robust}=0$.}
      \label{fig1a}
  \end{figure*}

The flux densities of the primary calibrators were set according to the  \cite{Perley2013} extension to the \cite{Baars1997}. Initial phase calibration was performed using primary calibrators and the solutions were subsequently used to compute parallel-hand delay. We performed standard bandpass calibration. Applying the bandpass and delay solutions, we proceeded to the gain solutions for the phase calibrators 3C286 and J0731+438. All relevant solution tables were applied to the target field. Each set was then averaged by a factor of 4 in frequency. 

After this, we combined all data. Several rounds of phase self-calibration were carried out on the combined data followed by two final rounds of amplitude and phase self-calibration. We visually inspected the self-calibration solutions and manually flagged some additional data. Deconvolution was performed with $\tt{nterms}$=2, $\tt{wprojplanes}$=500, using $\tt{Briggs}$ weighting with robust parameter 0. 

 \subsection{LOFAR-HBA observations}

The cluster was observed as part of the LOFAR Two-metre Sky Survey \citep[LoTSS;][]{Shimwell2017}. The  MACS\,J0717.5$+$3745 field was covered by one LoTSS pointing (P110+39) on 2019 December 5. For detailed description of the data reduction procedure, we refer readers to Rajpurohit et al submitted. 

To summarize, data reduction and calibration was performed with the LoTSS DR2 pipeline \citep{Tasse2020} The pipeline comprises a direction-independent calibration using \verb|PREFACTOR| \citep{vanWeeren2016c, Williams2016, deGasperin2019} and \verb|KillMS| \citep{Tasse2014a, Tasse2014b, Smirnov2015}. To create images of the entire LOFAR field-of-view, we used \verb|DDFacet| \citep{Tasse2018}. 

On the direction-dependent calibrated data, we performed an additional ``extraction + self-calibration'' scheme \citep{vanWeeren2020}. This method consists of subtraction of all sources, using the models derived from the pipeline, outside the cluster region. Afterwards, in the extracted region several rounds of phase and amplitude self-calibration were performed to optimize the solutions. In this process, the LOFAR station beam correction, at the position of the target, was also applied. We correct the flux density scale to NVSS to 6C sources \cite{Roger1973} flux densities (Shimwell et al. in preparation).

The uncertainty in the flux density measurements was estimated as: 
\begin{equation}
\Delta S =  \sqrt {(f \cdot S)^{2}+{N}_{{\rm{ beams}}}\ (\sigma_{{\rm{rms}}})^{2}},
\end{equation}
where $f$ is an absolute flux density calibration uncertainty, $S_\nu$ is the flux density, $\sigma_{{\rm{ rms}}}$ is the RMS noise and $N_{{\rm{beams}}}$ is the number of beams. We assume absolute flux density uncertainties 10\,\% for uGMRT \citep{Chandra2004} and LOFAR data.


\section{Results: Radio continuum images}
\label{results}
Our deep, ${\mathbf{\rm 8\arcsec\times\rm 4\arcsec}}$ resolution combined uGMRT band 3 and band 4 (300-850\,MHz) image of the cluster, with {\tt robust} =0 is shown in the left panel of Fig.\,\ref{fig1}. The uGMRT observation allows us to reach a sensitivity that is 5 times deeper with respect to the data published so far at this frequency \citep{vanWeeren2009,Bonafede2018,PandeyPommier2013}. 

The known diffuse emission sources, namely the chair-shaped relic (R) and the halo (H), are recovered in both uGMRT band\,3 and band\,4 observations; see Fig.\,\ref{fig1} and \ref{fig1a}. The sources are labeled following \cite{vanWeeren2017b} and extending the list.  An overview of the properties of the diffuse radio sources in the cluster is given in Table\,\ref{Tabel:Tabel2}. 

We measure a largest linear size (LLS) for the relic of $\sim$ 850\,kpc, which is in line with previous studies. The high-frequency VLA observations (L, S, and C-band) of the relic revealed several filaments on scales down to $\sim \rm 30 \, kpc$ \citep{vanWeeren2017b}. Some of these filaments originate from the relic itself, while a few of them appear more isolated and are located in the cluster outskirts. The uGMRT image also shows most of the structures reported at higher frequencies. To facilitate the discussion, we label some of the distinct features in the left and right panels of Fig.\,\ref{fig1}. 

One prominent feature, located to the north-west of R1, is the filament F1 (reported by \cite{vanWeeren2017b}, which is about 480 kpc. Another $\sim200$ kpc long filament, F2, connects the main relic to the halo.

In order to better separate fine structure across the relic from the surrounding diffuse, low surface brightness emission, we create an image with uniform weighting, see the right panel of Fig.\,\ref{fig1}. Its resulting image has a restoring beam of ${\mathbf{\rm 3.5\arcsec\times\rm 3.5\arcsec}}$, thus making structure stand out more clearly. The image shows that the northern part of the relic is composed of fine filaments (shown with red arrows). It remains unclear whether these filaments trace sites where particle are (re)accelerated similar to radio relics.

The radio halo emission is also well recovered with the uGMRT; see Fig.\,\ref{fig1a}. The halo extends to the north and south of R3 and R4. The detailed study of the halo emission will be presented in a subsequent paper (Rajpurohit et al.  submitted).

\begin{figure*}
    \centering
    \includegraphics[width = 0.99 \textwidth]{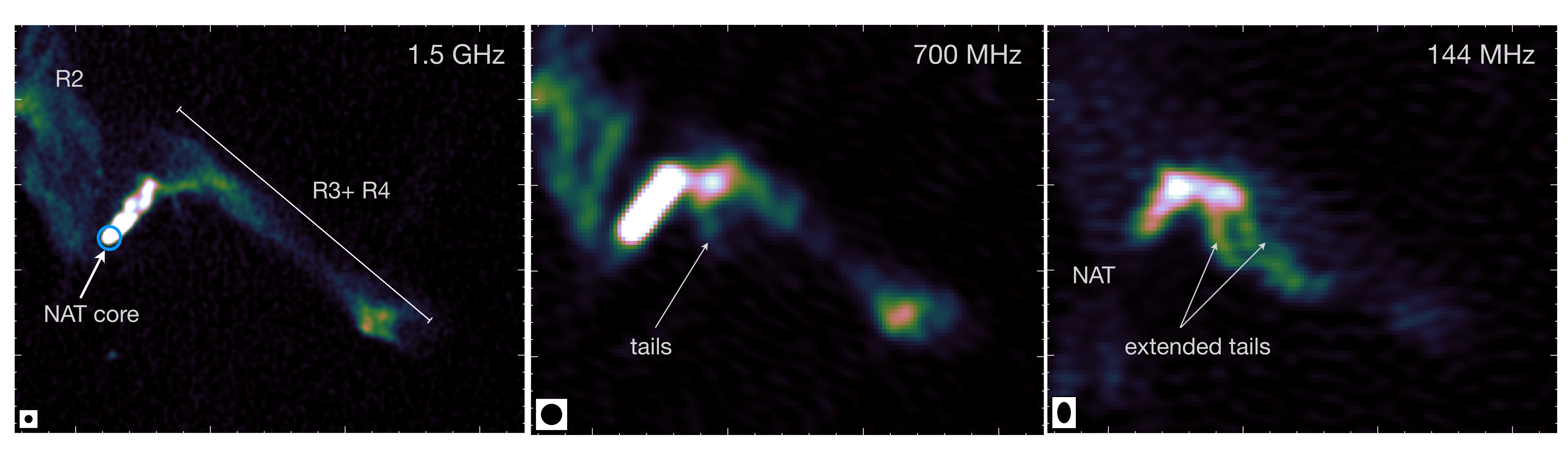}
    \caption{NAT galaxy located at the center of the relic at 1.5\,GHz (at $1.2\arcsec$ resolution), 700\,MHz (at $3.5\arcsec$ resolution), and 144\,MHz (at $4.5\arcsec\times3.4\arcsec$ resolution).  To compare radio  morphology of the NAT galaxy at different frequencies, the colors in these images were scaled manually. The image obtained with the new data evidently shows that both tails of the NAT bend down south rather fading or merging into the R3 region of the relic. The bent parts of tails are marginally visible in the VLA L-band image.  As discussed in Sect.\,\ref{nat}, the NAT and the southern part of the relic could be two different structures along the same line of sight seen in projection.}
    \label{fig::NAT}
\end{figure*}

The uGMRT 300-500\,MHz image is shown in the right panel of Fig.\,\ref{fig1a}. The morphology and extent of the relic similar to that of the uGMRT band\,4 images, however the halo emission is more extended. \cite{Bonafede2018} reported a large arc-shaped feature to the north-west of the relic at 147\,MHz which is tentatively classified as a relic. We do not detect the radio-arc in the observed uGMRT frequency range. To the south-east of the X-ray bar, a radio bridge has been reported connecting the radio halo to a head-tail radio galaxy, see Fig.\,\ref{XrayR} for the location of the galaxy \citep{Bonafede2018}. We do not detect the bridge between 300-850\,MHz, indicating that it thus must have a steep spectrum.

At the center of the main relic; see the right panel of Fig.\,\ref{fig1}, there is an embedded NAT galaxy with its tails aligned with the relic. At frequencies above 1\,GHz, the tails of the NAT are resolved and appear to fade into the R3 region of the relic; see the left panel of Fig.\,\ref{fig::NAT}. At 550-850 MHz, there is a hint of faint emission at the NAT that seems to bend to the south (middle panel). Interestingly, in the new LOFAR observation, the tails of the NAT are apparently bent to the south of R3; see the right panel of Fig.\,\ref{fig::NAT}. This suggests that the NAT is not necessarily morphologically connected to the R3 region of the relic but lays along the line of sight.

\section{Spectral analysis }
\label{spectral}

To study the spectral characteristics of the relic over a wide range of frequencies, we combine our uGMRT (300-850\,MHz) observations with those previously published VLA 1-6.5\,GHz \citep{vanWeeren2017b}. We also use the new LOFAR-HBA observations centered on 144\,MHz. For the LOFAR data reduction steps, we refer to Rajpurohit et al. (to be submitted). 

To accurately measure flux densities from radio interferometric images at different frequencies, it is necessary to ensure that all images reflect structures over the same range of angular scales. We achieve that here by imaging the data with uniform weighting scheme and a common lower uv-cut at $\rm 0.2\,k\uplambda$. Here, $\rm 0.2\,k\uplambda$ is the shortest, well-sampled baseline of the uGMRT data. This uv-cut is applied to the VLA L-band, S-band, C-band, and new LOFAR data. All maps were convolved to similar restoring beams, i.e., $8\arcsec$. We measure the flux density of the entire relic, as well as in four subareas R1, R2, R3, and R4 from the 144\,MHz to the 5.5\,GHz observations. The regions where the flux densities were extracted are indicated in the right panel of Fig.\,\ref{fig2}.

\subsection{Integrated spectrum}
\label{spectrum}
\begin{figure*}[!thbp]
\centering
 \includegraphics[width=0.99\textwidth]{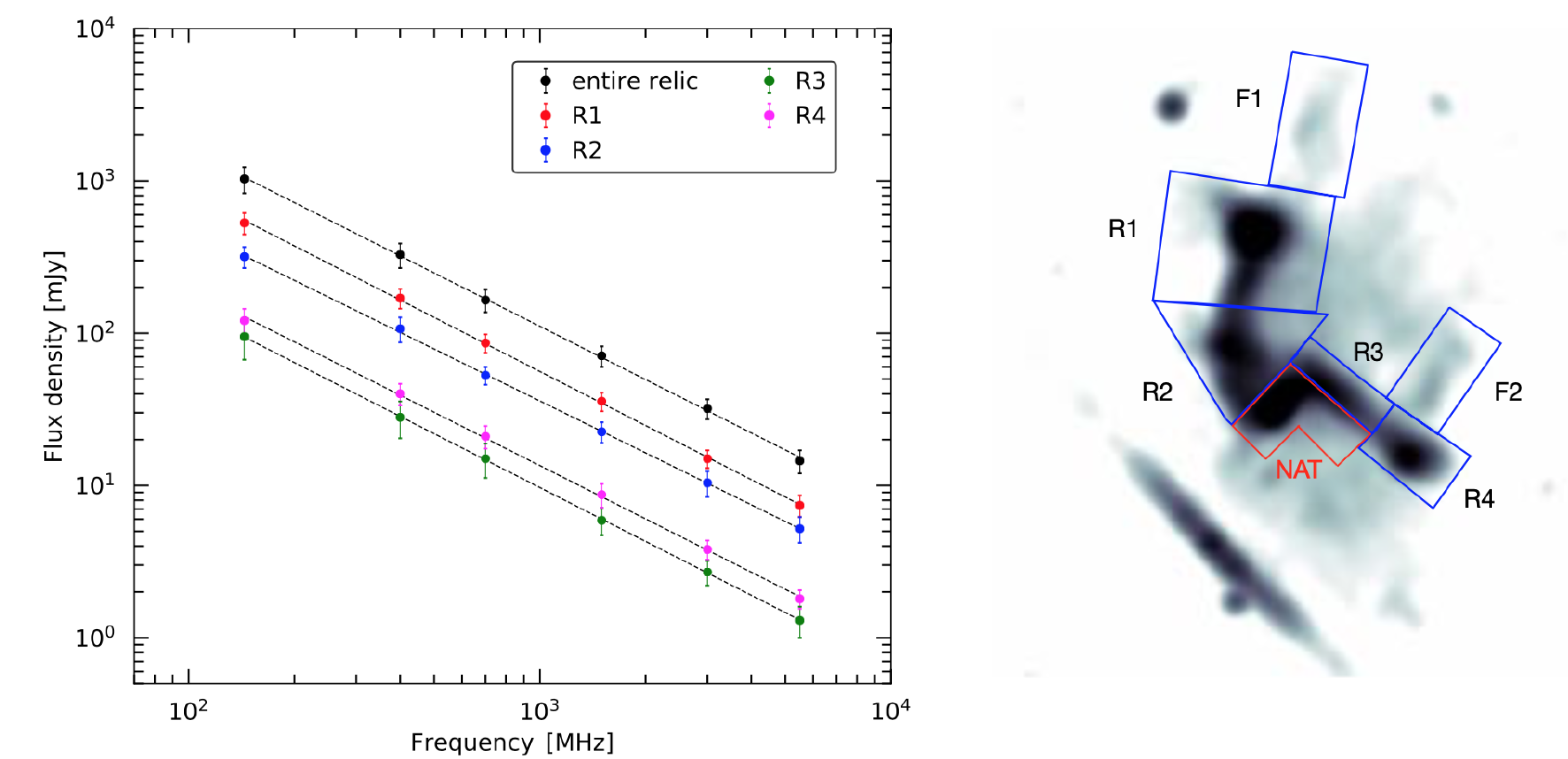}
 \vspace{0.0cm}
 \caption{{\it Left}: Integrated spectra of the chair-shaped relic and its subregions from 144\,MHz to 5.5\,GHz. The overall spectrum of the entire relic is well-described by a single power-law with the slope of  $\alpha=-1.16\pm0.03$. The subregions also follow a power-law distribution and show similar spectral slopes. As explained in Sec.\,\ref{spectrum}, the shock with Mach numbers ${\mathcal{M}}\sim3.5-4.0$ governs the integrated radio spectral index of relics. The total flux density of the relic and R3 is obtained after masking the contribution from the NAT (shown with red box). {\it Right}: Blue regions indicate areas from which the flux densities were extracted to obtain the integrated spectra of the relic and its subregions.}
 \label{fig2}
  \end{figure*}

The integrated synchrotron spectra of the entire relic and subregions, obtained by our  flux density measurements at frequencies 144\,MHz, 700\,MHz, and 1.5\,GHz, 3\,GHz, and 5.5\,GHz are shown in the left panel of Fig.\,\ref{fig2}. We measure a flux density for the entire relic as well as in sub-regions. The subregions boundary were set based on the high resolution L-band image. The NAT contamination of the total relic flux density was avoided by masking the NAT region; red box in the right Fig.\,\ref{fig2}. We find that a single power-law adequately fits the data up to 5.5\,GHz without any obvious break. The integrated relic emission between 144\,MHz and 5.5\,GHz has a spectral index of  $-1.16\pm0.03$.

DSA model for a shock of Mach number ${\mathcal{M}}$ makes two predictions for the synchrotron spectra of the accelerated electrons \citep{Blandford1987}. Immediately past the shock, the ``injected" electrons will produce a power law radio spectrum with spectral index $\alpha_{\rm inj}$, given by
\begin{equation}
\alpha_{\rm inj}=- \frac{1}{2} \left(\frac{1 + 3/{\mathcal{M}}^2}{1 - 1 / {\mathcal{M}^2}}\right).
\label{JEeq::alpha_inj_mach}
\end{equation}
Downstream from the shock, radiative cooling of the electrons results in an ever-steepening electron spectrum; 
when integrated over a distance larger than the electron cooling time, the spectral index of the ``integrated" power law spectrum will be
\begin{equation}
\alpha_{\rm int} =  \alpha_{\rm inj} -0.5  = -  \left(\frac{1 + 1 / {\mathcal{M}}^2}{1 - 1 / {\mathcal{M}^2}} \right).
\label{JEeq::alpha_int_mach}
\end{equation}
This integrated spectrum is the sum of regions with very different plasma ages, i.e., very different times since the shock has passed the emitting region.

\setlength{\tabcolsep}{8pt}
\begin{table*}[!htbp]
\caption{Properties of the diffuse radio emission in the cluster MACS\,J0717.5+3745.}
\centering
\begin{threeparttable} 
\begin{tabular}{ c | c | c | c | c | c | c | c | c }
\hline \hline
\multirow{1}{*}{Source} & \multicolumn{3}{c|}{VLA (1-6.5\,GHz)} & \multicolumn{2}{c|}{uGMRT (300-850\,MHz)} & \multirow{1}{*}{LOFAR} &\multirow{1}{*}{$\rm LLS^{\dagger}$} & \multirow{1}{*}{$\alpha_{144\rm \,MHz}^{5.5\,\rm GHz}$}$^{\dagger\dagger}$ \\
 \cline{2-6} 
&C-band & S-band & L-band  & band 4 & band3&  &  &\\
  \cline{2-6} 
&$S_{\rm 5.5\,GHz}$ & $S_{\rm3.0\,GHz}$ & $S_{\rm1.5\,GHz}$&$S_{\rm700\,MHz}$&$S_{\rm400\,MHz}$&$S_{\rm144\,MHz}$&&\\

 &mJy & mJy & mJy & mJy & mJy& mJy  &Mpc & \\
  \cline{2-3} \cline{3-5}\cline{6-7}
  \hline  
relic\,R &$15\pm3$& $32\pm5$ & $71\pm11$&$165\pm28$&$328\pm60$& $1030\pm197$&$\sim0.8$ &$-1.16\pm0.03$ \\ 
F1 &$0.31\pm0.01$ &$0.81\pm0.02$ &$1.6\pm0.3$&$4.0\pm1.0$&$7.0\pm3$& $22\pm4$ & $\sim0.5$&$-1.17\pm0.05$\\ 
F2 &$0.35\pm0.01$ &$0.70\pm0.02$ &$1.5\pm0.1$&$3.2\pm0.8$&$10.4\pm3$& $26\pm5$ & $\sim0.2$&$-1.18\pm0.05$\\ 
\hline 
\end{tabular}
\begin{tablenotes}[flushleft]
\footnotesize
\item{\textbf{Notes.}} Flux densities were extracted from $8.0\arcsec$ resolution images, created with uniform weighting and a uv-cut. The flux densities reported here are measured after masking the contribution from the NAT. The regions, where the flux densities were extracted, are indicated in the right panel of Fig.\,\ref{fig2}. We assume an absolute flux density scale uncertainty of 10\% for the uGMRT and LOFAR, 4\% for the VLA L-band, and 2.5\% for the VLA S-, C-band data. $^{\dagger}$the largest linear size at 700\,MHz; $^{\dagger\dagger}$ the integrated spectral index obtained by a single power-law fit. 
\end{tablenotes}
\end{threeparttable} 
\label{Tabel:Tabel2}   
\end{table*}

When the shock becomes strong -- ${\mathcal{M}} \gtw 3.5 $ or so -- the density jump across the shock depends only
weakly on $\mathcal{M}$. Because the density jump controls
the synchrotron spectral index, both $\alpha_{\rm inj}$ and $\alpha_{\rm int}$ also depend only weakly on $\mathcal{M}$.  For instance, between
$\mathcal{M} = 3.5$ and $\mathcal{M} = 4.0$, the integrated spectral index varies only between $\alpha_{\rm int} \simeq -1.18$ to $-1.13$. 
 
The integrated spectral index of the whole relic, $\alpha_{\rm int}= -1.16\pm 0.03$, corresponds to a shock of Mach number $\mathcal{M}=3.7\pm0.3$.
Furthermore, as seen in the left panel of Fig.\,\ref{fig2}, the subregions of the relic also follow a power-law distribution, despite their complex morphology. The mean spectral indices of regions R1, R2, R3, and R4 are $-1.18\pm0.03$, $-1.13\pm0.03$, $-1.17\pm0.05$, and $-1.16\pm0.03$, respectively. Interpreting these values as integrated spectra from DSA, the subregions are dominated by shocks with ${\mathcal{M}}\simeq3.5-4.0$. Recently, \citet{Rajpurohit2020a,Rajpurohit2020b} investigated the integrated spectrum of the Toothbrush relic. They found that the entire relic and subregions follow a closely power-law between 58\,MHz to 18.6\,GHz and show almost identical slopes.  

As argued in \cite{Rajpurohit2020a}, the dependence of the radio luminosity on the Mach number determines the spectral index of the radio emission in relics. The shock front shows a distribution of Mach numbers but a single Mach number derived above can only provide a first-order characterization of the shock. More specifically, the Mach numbers derived from the radio integrated spectra are dominated by the tail of the high Mach number shocks \citep{2019arXiv190911329W,Rajpurohit2020a,Paola2020}. It is worth to emphasize here that the MACS\,J0717.5+3745 and the Toothbrush relics show, for all regions, a ``universal" spectral index of about $-1.16$. In addition, the simulated relic discussed in Sect.\,\ref{simlation_part} also gives an almost identical Mach number;  see also  \cite{Rajpurohit2020a}.

\begin{figure*}[!thbp]
    \centering
    \includegraphics[width=0.99\textwidth]{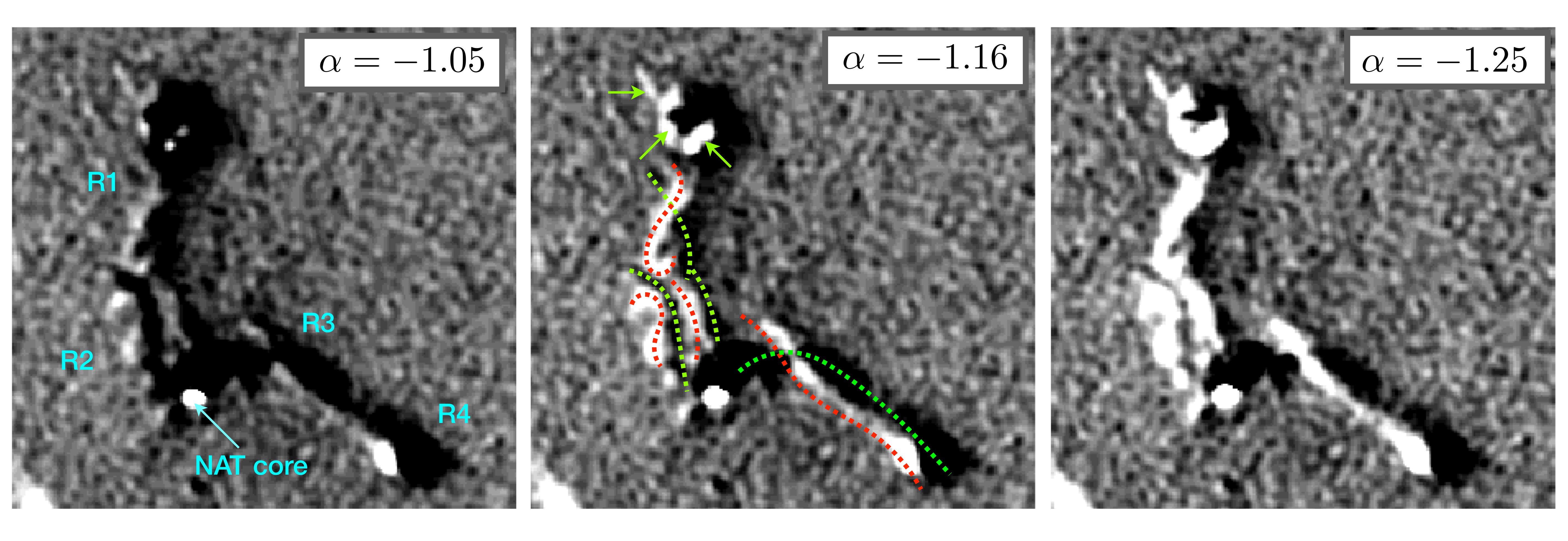}
    \vspace{-0.0cm}
    \caption{Gallery of spectral tomography maps between 1.5 and 5.5\,GHz at 2.5 \arcsec resolution. The images demonstrate that there are structures with different spectral indices overlapping with each other (shown with red and green dashed lines). Red lines in each frame indicate spectra which are  flatter than the listed $\alpha_{i}$, while green lines indicate spectra steeper than the $\alpha_{i}$. This implies that the relic is composed of multiple overlapping structures. The small scale structures are shown with green arrows.  The emission with a spectrum steeper than $\alpha_{i}$ appears positive (black regions), while flatter spectrum emission appears negative (white regions). The range in $\alpha_{i}$ is -1.05, -1.16, and -1.25.}
      \label{ST}
\end{figure*}

This raises the question why both in observations and simulations, relics show the same broadband power-law spectra and suggest a Mach number of $\mathcal{M}\sim 3.7$. We argue this comes from the density jump across the shock being very insensitive to shocks with Mach number $\gtw 3.5$, 
as discussed above. It follows that neither the integrated or injection index change much when the shock Mach number gets large. Hence, the regions with large Mach number along the shock front appear to dominate the relic emission.

F1 and F2 also follow a power-law distribution. The integrated spectral indices of filaments F1 and F2 are $-1.17\pm0.05$ and $-1.18\pm0.05$, respectively. In high resolution VLA images, F1 and F2 are apparently isolated from the relic. Similar distinct, isolated and diffuse filaments are found for relic in CIZA\,J2242+5301 \cite[hereafter ``Sausage relic";][]{Gennaro2018}. The presence of such distinct features may indicate the distributions of shock(s) caused by the same merger event. The complex ICM distribution causes differences in the sound speed and pressure gradients which ultimately affect the propagation of the shock. Therefore, it is plausible to assume that a single merger event can cause a complex shock front distribution.

 \begin{figure*}[!thbp]
    \centering
         \includegraphics[width=0.49\textwidth]{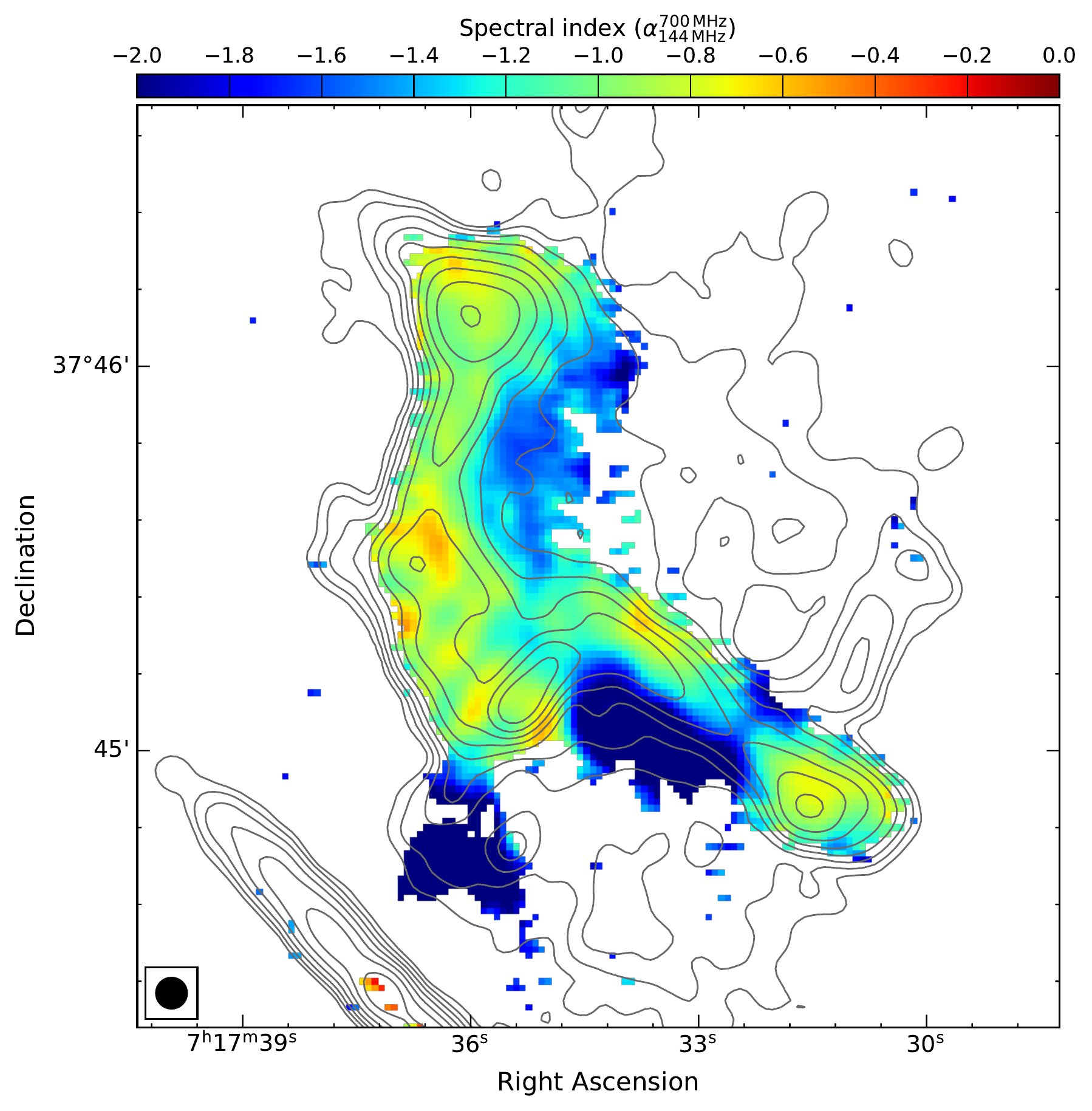}
     \includegraphics[width=0.49\textwidth]{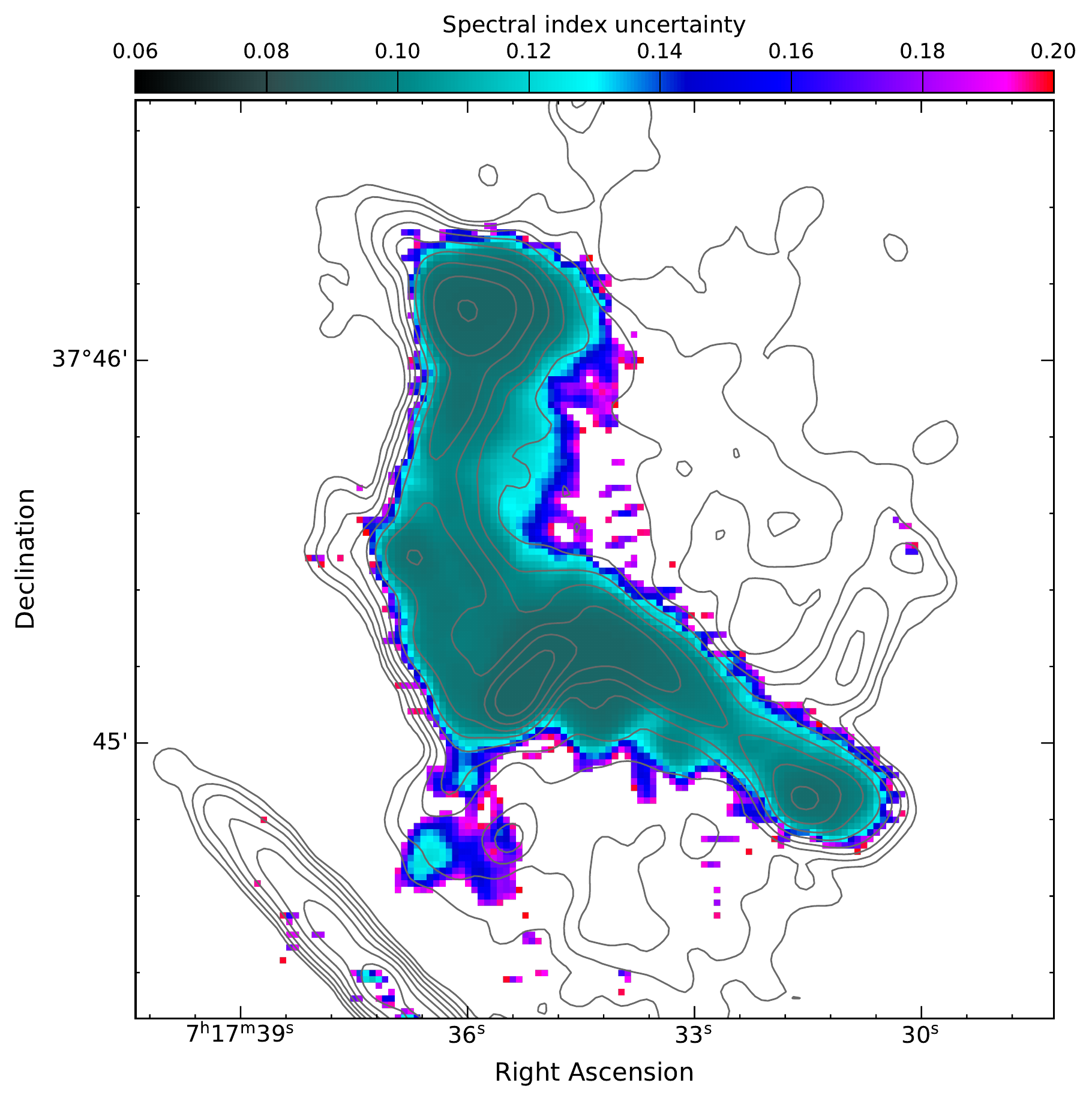}
         \includegraphics[width=0.49\textwidth]{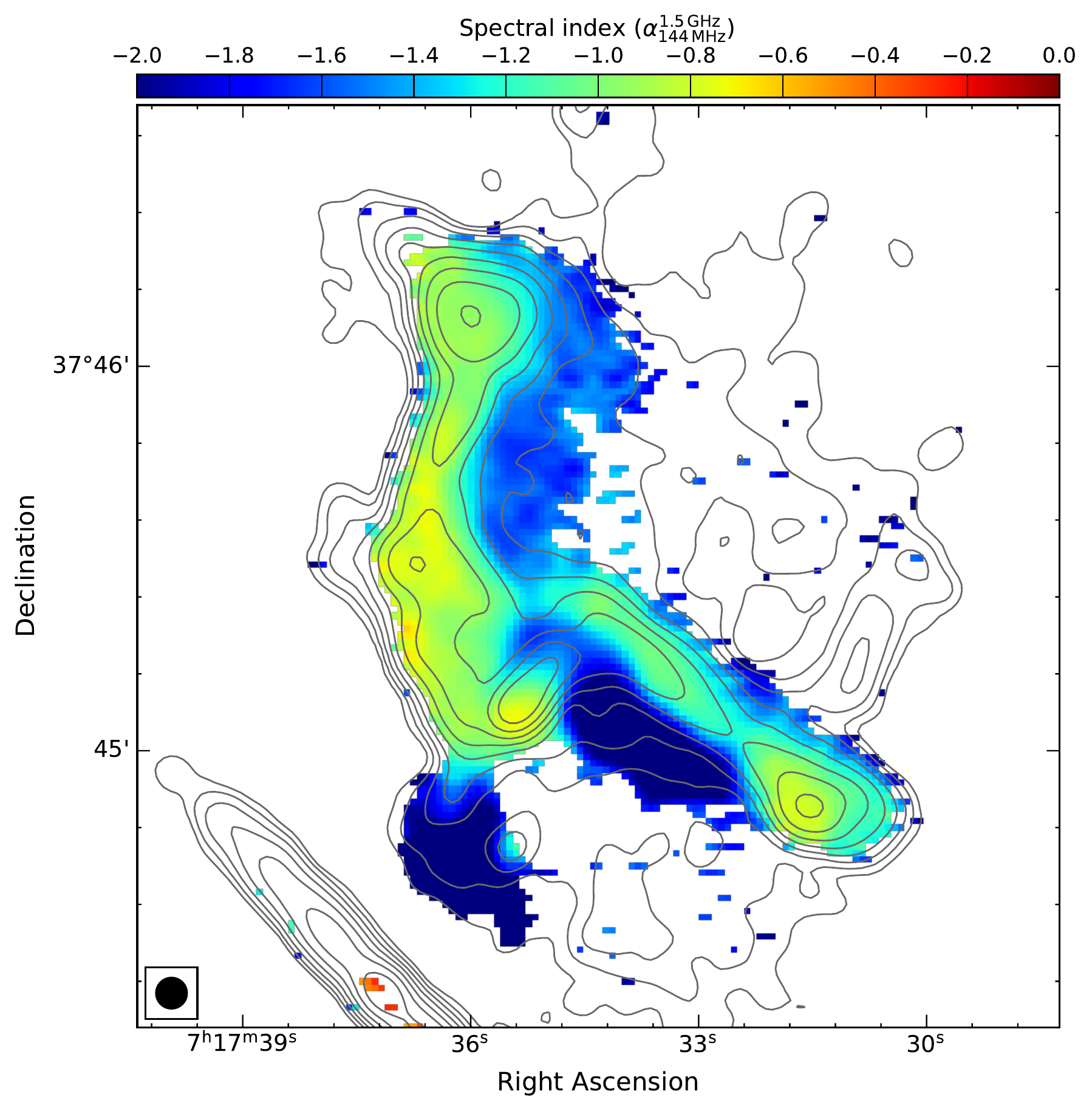}   
      \includegraphics[width=0.49\textwidth]{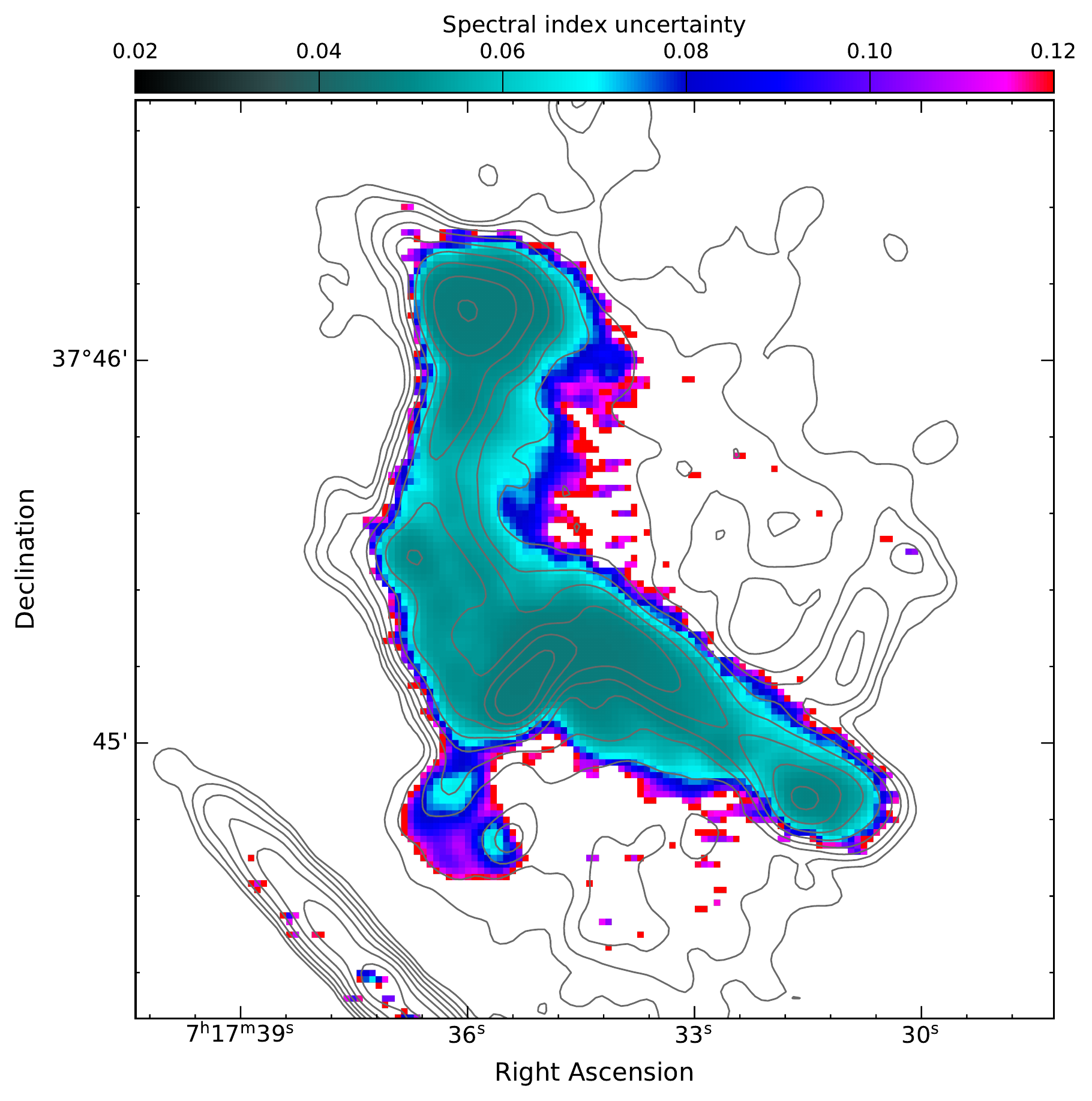}
    \vspace{-0.3cm}
    \caption{\textit{Left}: Spectral index maps of the relic in MACS\,J0717.5+3745 at $4.8\arcsec$ resolution from 144-700\,MHz (top) and 144\,MHz -1.5\,GHz (bottom). The injection spectral index at the outer edge along the relic varies between $-0.70$ to $-0.90$, reflecting inhomogeneities in the injection index. The color bar shows spectral index $\alpha$ from $-2.5$ to 0. Contour levels are drawn at $\sqrt{[1,2,4,8,\dots]}\,\times\,4.5\sigma_{{\rm{ rms}}}$ and are from the VLA L-band image. The beam sizes are indicated in the bottom left corner of the each image. \textit{Right}: corresponding spectral index uncertainty maps.}
      \label{SI}
\end{figure*}

\begin{figure*}[!thbp]
    \centering
    \includegraphics[width=0.49\textwidth]{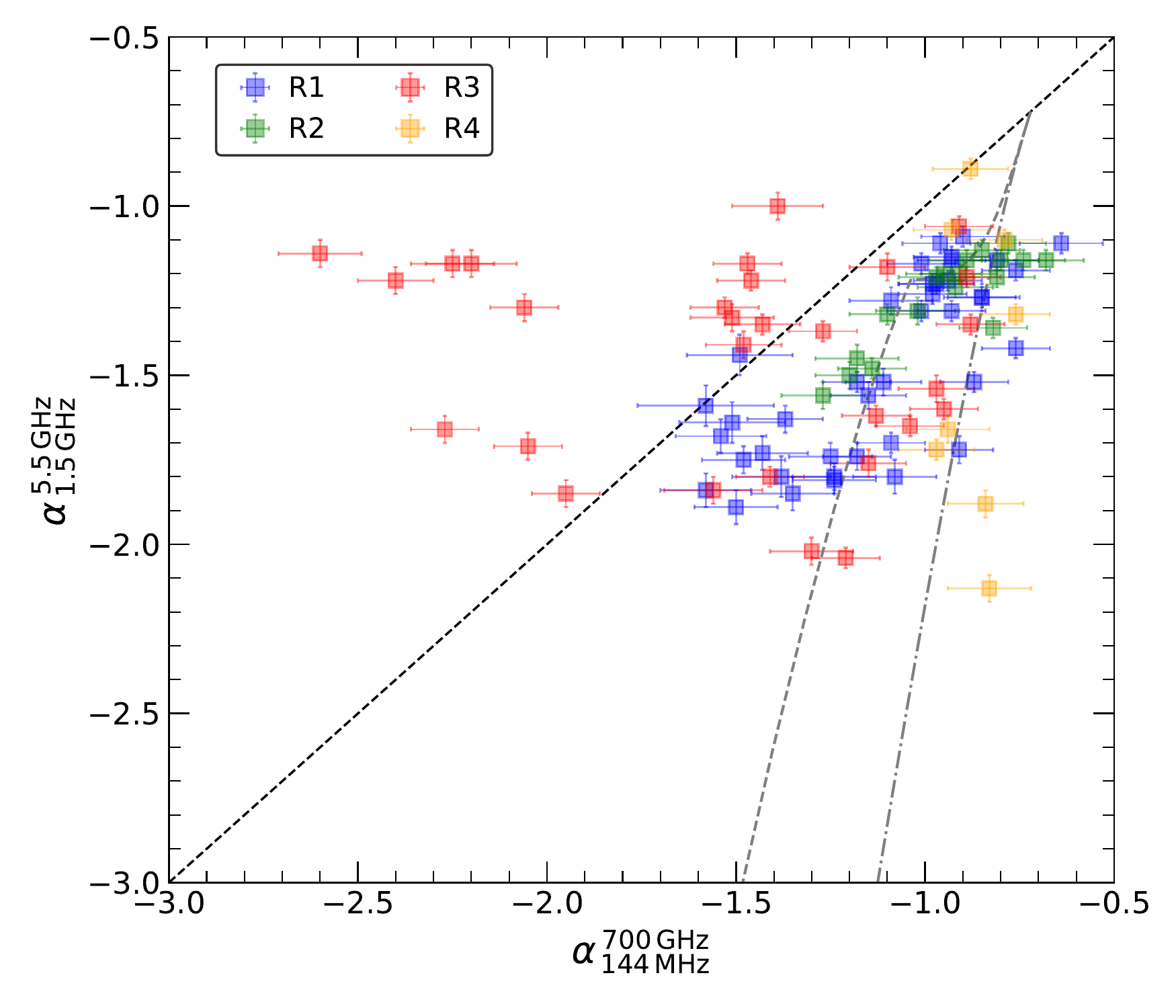}
       \includegraphics[width=0.49\textwidth]{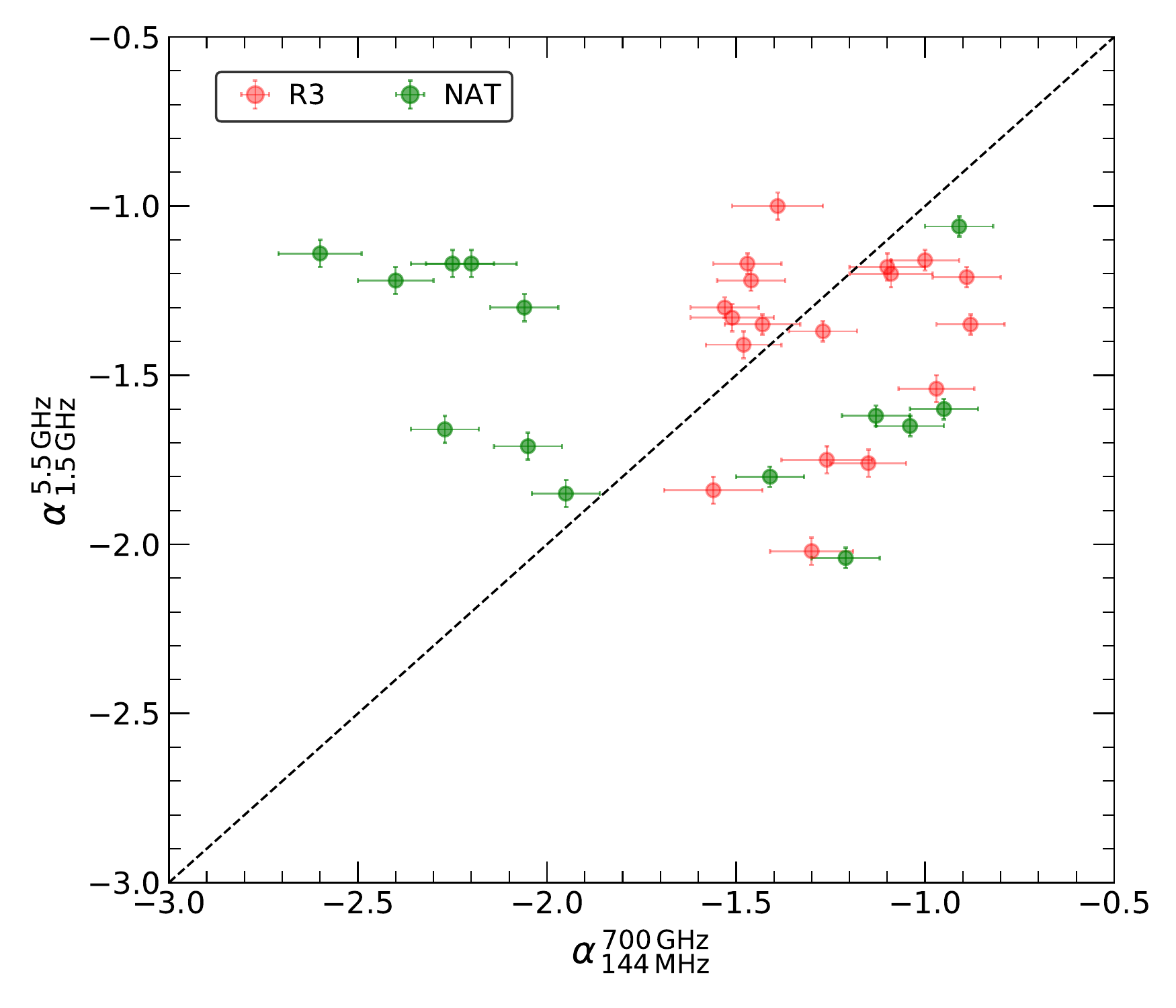} 
 \vspace{-0.2cm}
    \caption{Radio color-color plots of the relic in MACS\,J0717.5+3745, superimposed with the JP (gray dash-dotted line) and KGJP (gray dashed line) spectral aging models obtained with $\alpha_{\rm inj}= -0.70$. \textit{Left}: Majority of the relic points are below the power law line, indicating negative spectral curvature. Many of the data points from the R3 region (red) of the relic lies in the concave part (positive curvature) of the color-color plot. The spectral curvature is not consistent with any of the standard spectral aging models. \textit{Right}: Color-color plot showing that the NAT and the R3 region of the relic have different spectral shapes, suggesting two overlapping components along the line of sight. To extract spectral index values, we create a grid of squared boxes with a width of 4.8\arcsec, corresponding to a physical size of about 31\,kpc.}
          \label{CC_plot}
\end{figure*}

\subsection{Spectral tomography: filaments overlapping}

To investigate if there are any local spectral index variations or overlapping features, we construct images using the ``spectral tomography'' technique, first introduced by \cite{KatzStone1997}. In this technique, scaled versions of one image are subtracted from another: 
\begin{equation}
I(\alpha_{i})=I_{\nu_{1}}-\left(\frac{\nu_{1}}{\nu_{2}}\right)^{\alpha_{i}} I_{\nu_{2}},
\end{equation} 
where  $\nu_{1} = 1.5\,\rm GHz$ and $\nu_{2} = 5.5\,\rm GHz$.  If a structure has a spectral index identical to $\alpha_{i}$, the structure will disappear. Spatial components with a spectral index flatter than $\alpha_{i}$ will be over-subtracted with respect to its surrounding emission and have negative brightness. If a structure has a spectral index steeper than $\alpha_{i}$, it will be under-subtracted and have positive  brightness. This technique allows us to trace fine-scale spectral index changes, and is particularly useful for disentangling the superposition of components with different spectra that can overlap along the line of sight. Since we do not know a priori for the spectral index  to remove a particular structures, we construct a gallery of maps.

In Fig.\,\ref{ST}, we show the resulting images for $\alpha_{i}= \rm-1.05,\,-1.16, and \,-1.25$.  In these maps, the emission with a spectrum steeper than the above respective values of $\alpha_{i}$ appears positive (dark regions), while flatter spectrum emission appears negative (light regions). For example, the core of the NAT is flatter in all maps and shows up white. 

High-resolution observations of relics provide evidence of filaments within the relics, for example in Abell\,2256 \citep{Owen2014}, the Toothbrush relic \cite{Rajpurohit2018,Rajpurohit2020a} and the Sausage relic \citep{Gennaro2018}. These filaments may partly reflect the variation of radio luminosity along the shock front, which is slightly inclined with respect to the line of sight or the distribution of magnetic fields in the radio-emitting volume. However, filaments may also caused by plasma instabilities due to the mixing of relativistic and thermal plasma and magnetic field loops where electrons are trapped. 

From the spectral tomography maps (Fig.\,\ref{ST}), it is evident that there are fine structures across the relic (shown with red and green lines). These structures have either steeper (green lines) or flatter (red) spectral indices than the corresponding $\alpha_{i}$. This implies that there are structure in the relic with different spectral indices overlapping with each other. At the northern part of the relic, some of these features cross each other and some runs parallel. The tomography images reveal that relic is likely composed of multiple fine structures. As visible in the spatially resolved spectral index maps \citep{vanWeeren2017b}, we can also see a strong spectral index gradient across the entire relic. At the location of the NAT and R3, there is evidence for two spectral components.

\subsection{Spectral index and curvature}
\label{indexmaps}

We investigate the spectral index and curvature across the relic in MACS\,J0717.5+3745. \citet{vanWeeren2017b} reported high frequency spectral index maps  of the relic. They found that the relic shows a clear spectral index gradient between 1.5 to 5.5\,GHz; from east to west for the northern part and south to north for the southern part. Radio relics typically show spectral index gradients towards the cluster center \citep{vanWeeren2010,Bonafede2012,Gennaro2018,Rajpurohit2020a}. This spectral steepening is believed to reflect the aging of relativistic electrons of an outward moving shock due to the combined effect of synchrotron and Inverse Compton losses.  Some studies show that the spectral index gradient should not necessarily be interpreted as the aging of electrons but could be evidence of varying Mach number across the shock surface \citep{Skillman2013,deGasperin2015}.

In \cite{Bonafede2018}, a low-frequency spectral index map of the relic was derived using the LOFAR 150\,MHz and GMRT 610\,MHz data. However, the resolution of the low-frequency image was limited to 10\arcsec. We aim to create spectral index maps between 144\,MHz to 1.5\,GHz at the highest possible common resolution, namely at $4.8\arcsec$ resolution. For the spectral index maps at two frequencies, we blank pixels below $4.5\sigma_{\rm rms}$. The resulting spectral index maps for two pairs of frequencies are shown in Fig.\,\ref{SI}. 

\begin{figure}[!thbp]
    \centering
    \includegraphics[width=0.42\textwidth]{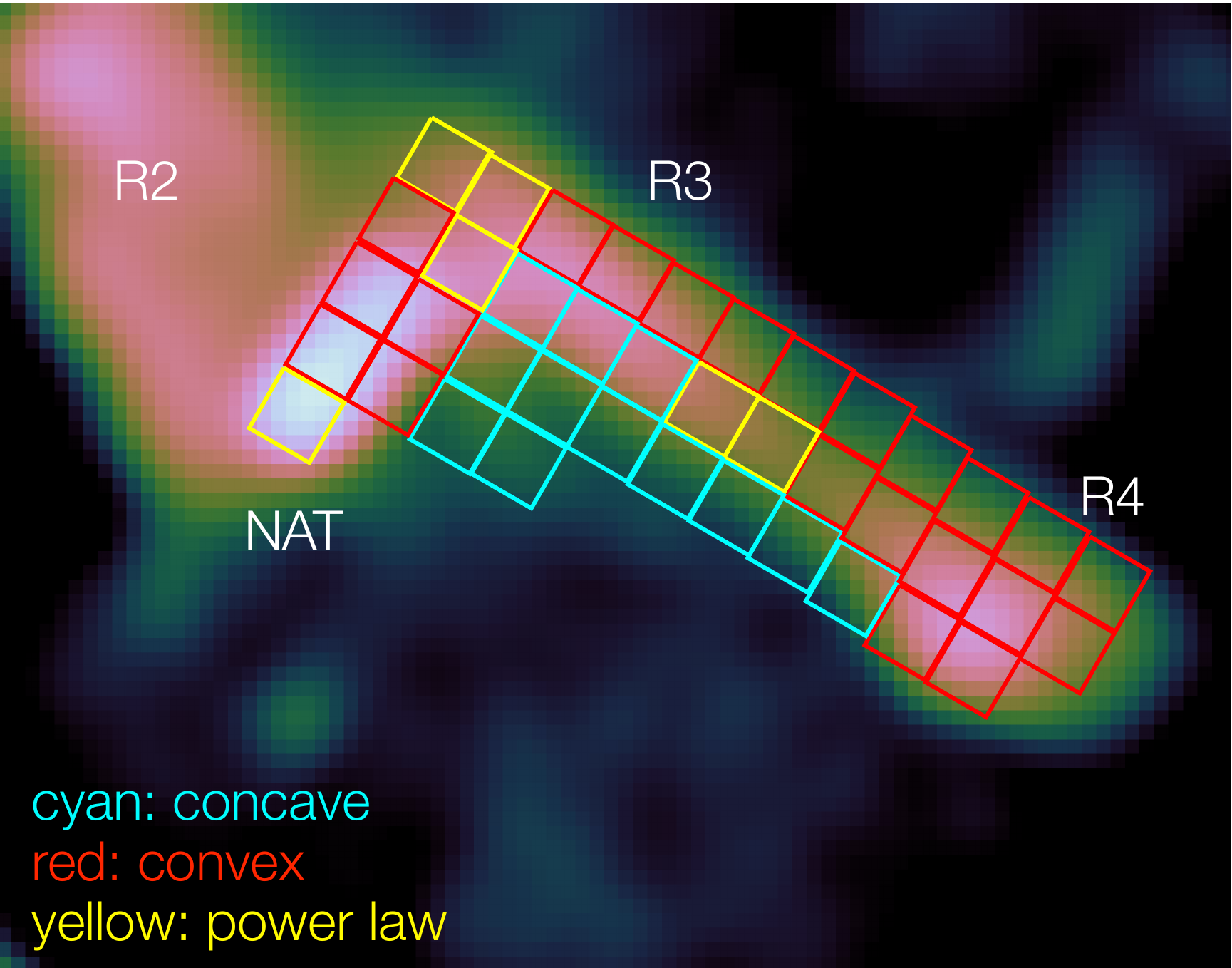}
    \caption{Spectra across the southern part of the relic and the NAT, overlaid on the L-band total intensity image. The different spectral shapes are evident. The long tails of the NAT show concave spectra, which could be due the superposition of two different electron populations along the same line of sight. }
          \label{nat_plot}
\end{figure}  

For the relic, the low-frequency spectral index values are flatter than those reported at higher frequencies \citep{vanWeeren2017b}. Between 144\,MHz to 1.5\,GHz, the spectral index across the relic mainly varies between $-0.70$ to $-1.80$. At high frequencies (1.5 to 5.5\,GHz), the spectral index steepens from about $-0.60$ to $-1.5$. The spectral trends for the northern part of the relic are similar to those reported at high frequencies. However, the spectral trends are different at the NAT galaxy and R3. The NAT shows a progressive steepening with increasing distance from the core, reaching values up to $-2.8$. This indicates that the steeper part, to the south of R3, is an extension of the NAT tails. The steepening is stronger in the low-frequency range than at higher frequencies \citep{vanWeeren2017b}.  In Section\,\ref{nat}, we suggest that the NAT and the southern part of the relic are not connected which is consistent with the spectral index properties.

The ``injection" spectral index at the shock, from Eq.\,\ref{JEeq::alpha_inj_mach},  can be measured from the outer edge of the relic where the acceleration process is presumed to happen. The low-frequency maps are better suited for estimating the injection index as the downstream profile is wider; see \citet{Rajpurohit2018,Rajpurohit2020a} for discussion. For example, at the eastern edge of the R1 and R2 regions, the flattest spectral index in the 144\,MHz to 1.5\,MHz map is clearly flatter than the one between 1.5-5.5\,GHz \citep{vanWeeren2017b}. We measure an injection index ranging from $-0.70$ to $-0.92$ between 144\,MHz to 1.5\,GHz. Unlike integrated spectra, the spatially resolved index maps show strong variation in the injection index. This possibly indicates inhomogeneities in Mach numbers across the shock or variation in the magnetic field strength. 

For the relic, we measure an injection index as flat as $-0.70\pm0.04$, corresponding to a Mach number of $\mathcal{M}=3.3\pm0.3$. However, the combination of projection effects and smoothing in the radio image may make it hard to measure the actual injection index. Therefore, the injection index measured from our spatially resolved spectral index map may be flatter than observed.

The Mach number obtained from the injection index is more or less consistent with the one obtained from the integrated spectrum, namely $ \mathcal{M}=3.7$. For the relic in MACS\,J0717.5+3745, we do not find any significant discrepancy between the Mach number derived from the integrated spectrum and the injection index. Indeed, radio observations imply a high Mach number shock. However, in deep Chandra observations, no X-ray surface brightness discontinuity has been detected at the location of the relic \citep{vanWeeren2017b}. The non-detection of a shock in X-ray observations suggests that the shock surface is not seen close to edge-on or that projection effects may hide the surface brightness jump. In Sect\,\ref{indexmaps}, we discuss this in further detail.

Merger-shock models predict increasing spectral curvature in the post-shock areas of relics, e.g., the Toothbrush and the Sausage relics \citep{vanWeeren2012a,Stroe2013,Gennaro2018,Rajpurohit2020a}. To determine whether the relic shows any sign of spectral curvature, we employ the color-color diagram method \citep{Rudnick1994,vanWeeren2012a,Rajpurohit2020a}. Since conditions in the relic could be inhomogeneous, we create maps at $4.8\arcsec$ resolution using the new LOFAR HBA, uGMRT, and the VLA L, S, and C-band data. This resolution minimizes the mixing within a single beam of emission, thus permitting the detection of features with different spectral properties. The images were created using  the same uv-cut and uniform weighting as described in Sect\,\ref{spectral}. The spectral indices were extracted from a grid of $4.8\arcsec$ boxes (i.e., the beam size) covering the relic. For low frequencies, we use our spectral index map created between 144 \,MHz and 700\,MHz while for the high frequencies between 1.5\,GHz and 5.5\,GHz. 

The resulting color-color plots are shown in Fig.\,\ref{CC_plot}. The color-color plots are similar to spectral curvature maps, with the curvature being the difference between the low and high spectral indices. The curvature is negative for a convex spectrum. We find that the majority of the relic points fall below the black dashed line ($\alpha_{144\,\rm MHz}^{700\,\rm MHz}=\alpha_{1.5\,\rm GHz}^{5.5\,\rm GHz}$), which is a clear signature of a negative curvature. Except for some of the points in R3, the different regions of the relic seems to follow a single locus. We note that the curvature at R4 seems too convex, indicating that R4 is possibly different.  

We overlay the color-color plot with the Jaffe-Perola \citep[JP; ][]{Jaffe1973} and KGJP \citep{Komissarov1994} spectral aging models. These models take into account the injection index and subsequent radiative losses. We find that none of the spectral aging models can fully explain the observed spectral shape. For the relic in MACS\,J0717.5+3745, the observed trajectory in the color-color is different than that of the Toothbrush and the Sausage relic. The physical origin of the shape and the curvature distribution in the present relic remains unclear. As discussed in Sect.\,\ref{simlation_part}, we speculate that the relic is likely to be inclined along the line-of-sight, and that projection effects might be responsible for the observed spectral shape.

\subsection{Shock re-acceleration by the NAT?}
\label{nat}

Some radio relics are believed  to be likely caused by re-acceleration of fossil electrons, previously injected by a nearby AGN, for example Abell 3411-3412 \citep{vanWeeren2017a} and CIZA\,J2242+5301 \citep{Gennaro2018}. It has been suggested that the southern part of the relic in MACS\,J0717.5+3745 has been powered by shock re-acceleration of fossil electrons, from the NAT \citep{vanWeeren2017b}. This claim was based on three major points: (1) the NAT is a cluster member: (2) a direct morphological connection between the relic and the NAT; (3) spectral flattening at the location where the NAT tails meet the southern part of relic.

\begin{figure}[!thbp]
    \centering
    \includegraphics[width=0.48\textwidth]{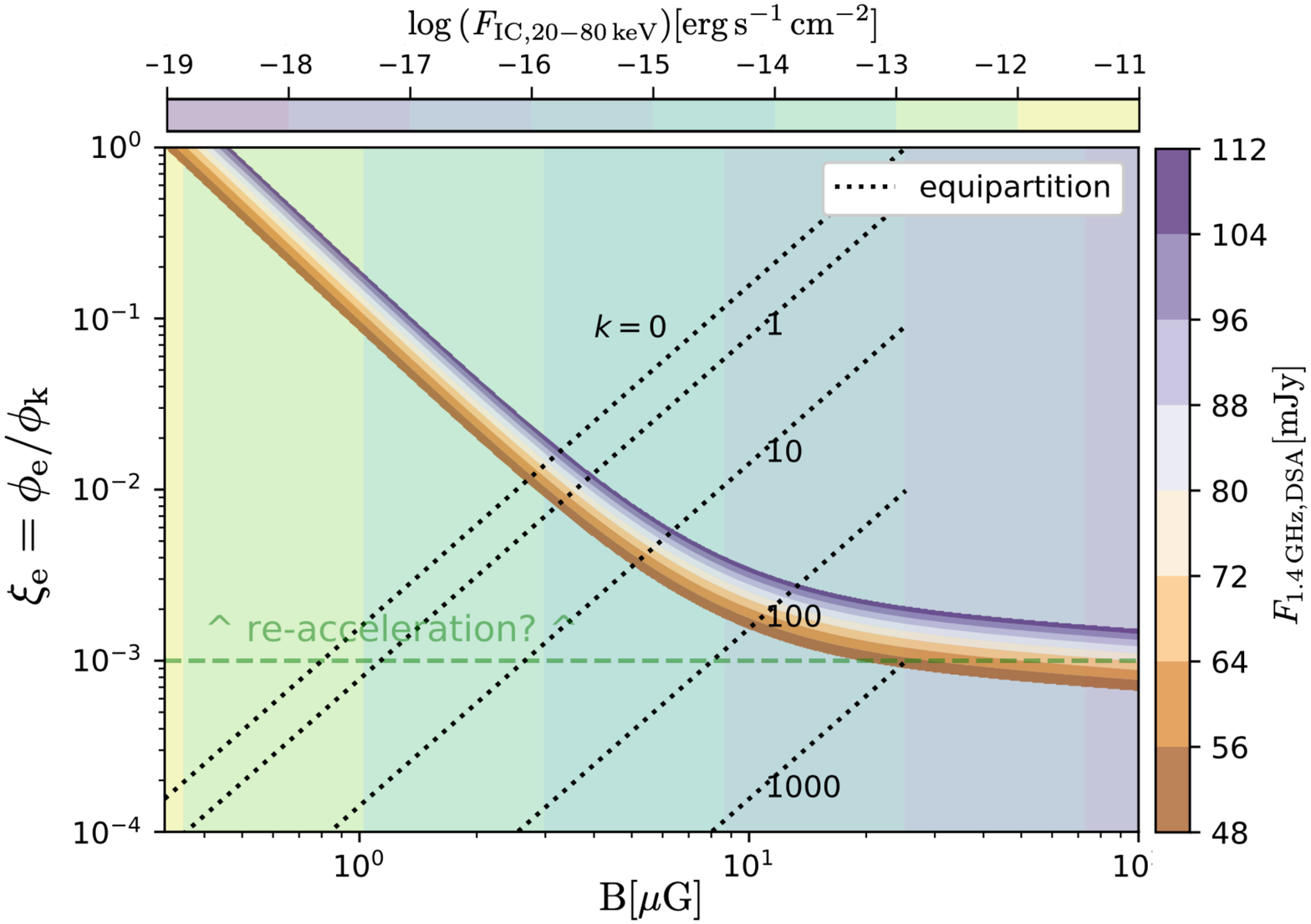}
    \vspace{-0.1cm}
    \caption{Electron acceleration efficiency ($\xi_e$) for the relic in MACS\,J0717.5+3745 versus magnetic field ($B$). Calculations were performed assuming the DSA scenario \citep{Hoeft2007} using $\alpha=-1.16$. The curve show the points that reproduce the $S_{1.4\,\rm GHz}$ flux density. Black dashed lines show the values obtained assuming equipartition for different values of $k$; see \cite{Locatelli2020} for details. The horizontal green dashed line shows the theoretical upper limit allowed by standard DSA, namely at $\xi_e=10^{-3}$. The background colors predict the inverse Compton flux expected in the 20-80~keV band, produced by the same electrons responsible for the relic emission. An acceleration efficiency below 0.01 results in the observed radio luminosity of the relic when the magnetic field is in the range of $B=1-6\,\upmu\rm G$.}
 \label{figxyz}
\end{figure}

\begin{figure*}
    \centering
    \includegraphics[width = 0.49\textwidth]{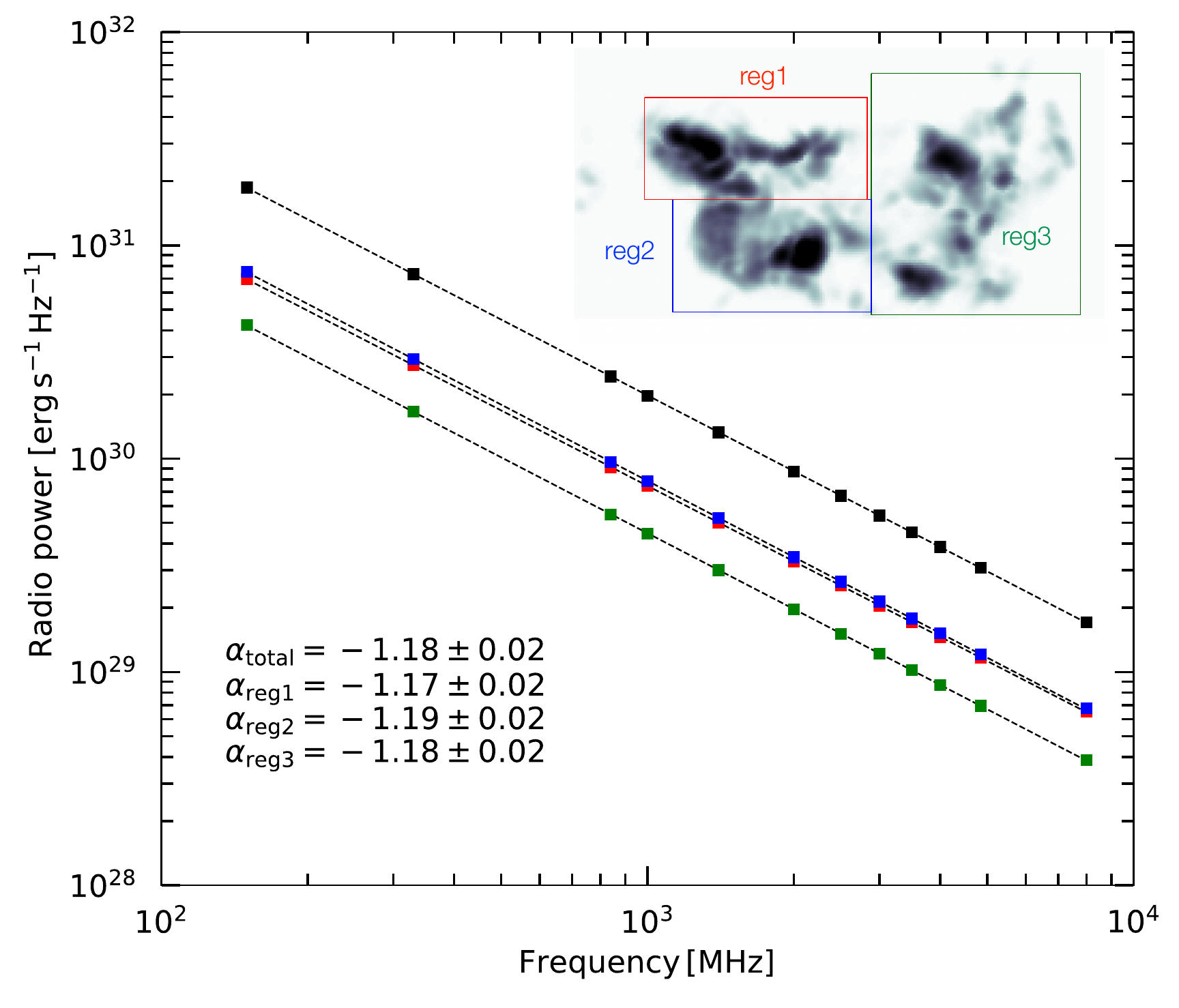}
        \includegraphics[width = 0.49\textwidth]{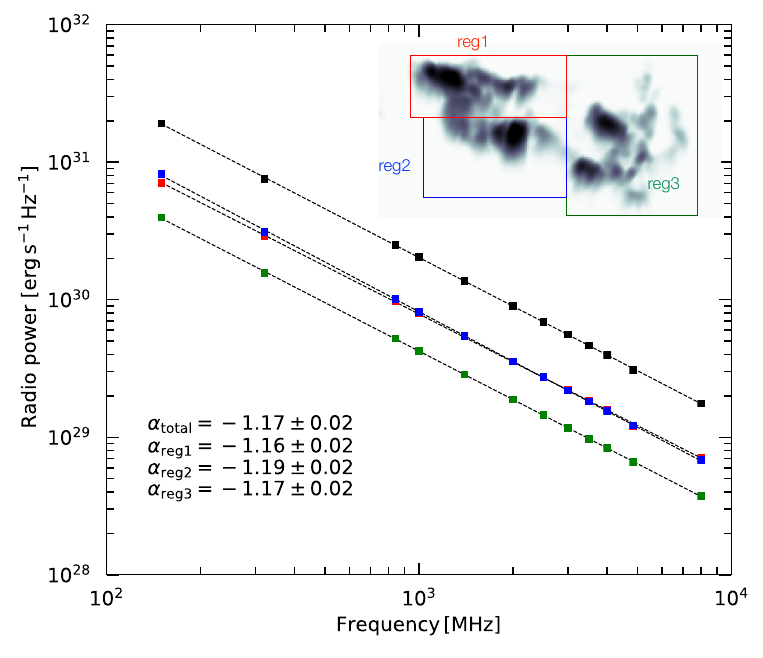}
    \vspace{-0.2cm}
    \caption{Integrated spectra of the relic when seen face-on (left) and at 45\degree (right), simulated in \citet{2019arXiv190911329W}. The black spectrum was measured across the whole relic, while the other spectra were taken in the different subregions as shown in the inset. The inset map gives the total intensity at $1.4 \ \mathrm{GHz}$. The simulated relic is $\sim 1.3 \ \mathrm{Mpc} \times 0.7 \ \mathrm{Mpc}$ in size. Like an edge relic, the face-on and the 45\degree rotated relics also show almost similar spectral slopes. This implies that the integrated radio spectrum of relic is independent of the viewing angle. For the integrated spectra of the relic seen edge-on, we refer to Fig. 8 in \citet{Rajpurohit2020a}.}
    \label{fig::simu_spectrum}
\end{figure*}

As visible in the right panel of Fig.\,\ref{CC_plot}, the R3 region of the relic shows a sample of points that lies above the power-law line. The concave spectra are only seen for R3 and nowhere else in the relic. These points represent a region with positive (concave) curvature, i.e., the spectrum is flatter at higher frequencies. This is the expected behavior in the case of a superposition of different emission spectra along the same line of sight. If some features exhibit different spectral indices, regions with flatter spectra will dominate at higher frequencies, resulting in the spatially concave spectrum. To investigate where exactly the concave regions are coming from, we color-coded the spectra; see Fig.\,\ref{nat_plot}. The concave spectra are mainly found at locations where tails of the NAT meet the relic (cyan boxes). 

The spectral index across the NAT (from the core to tails) gradually steepens from $-0.67$ to $-2.8$ between 144\,MHz and 700\,MHz . In contrast, between 1.5 and 5.5\,GHz, it is consistently flatter, and changes from $-0.60$ to $-2.3$, opposite to what is expected for an aging spectrum. From the total intensity maps, it is evident that at the NAT there is also a faint relic component. This suggests that emission from both the NAT and the relic are present, albeit in different proportions at different frequencies. Since the bent part of the NAT tails is marginally visible above 700\,MHz, at low frequencies the emission from the NAT dominates that of the relic emission and the spectral index gradually steepens. Such a spectral steepening is expected for radio galaxies when electrons suffer energy losses. In contrast, at high frequencies the bent tails are marginally visible, thus relic emission dominates and the spectral index remains relatively flatter. 

Based on the new observations we raise the possibility that R3 and the NAT are two different structures projected along the line of sight. This suggests that that the relic is not powered by seed electrons from the NAT. In detail: (1) our new broadband data shows the tails of the NAT bend to the south, rather than merging or fading into the R3 region of the relic (see Fig. \,\ref{fig::NAT}); (2) the spectral index at the NAT is different from the relic spectral index: a clear spectral steepening from the core to tails is instead seen across the NAT, indicating that the steeper regions are the extension of the NAT tails; (3) the concave spectra at the location of the NAT and R3 seems compatible to the superposition of increasing steep spectra from the NAT and the flatter spectra from R3. In addition, the spectral tomography also reveals the presence of two spectral components at the NAT and R3. In summary, the morphology and spectral index images raise the question about whether the two structures are merely overlapping or physically interacting. A further discussion of this issue will be a presented in a subsequent paper (Rajpurohit et al. to be submitted) containing polarization. We note that the reacceleration of ``ghost'' plasma (NOT connected with the NAT) is still possible, however we do not explore this possibility in detail and focus on acceleration of the electrons by the DSA form the thermal pool.


\section{Acceleration efficiency and magnetic fields}
\label{eff}

\begin{figure}
\centering
\includegraphics[width = 0.45 \textwidth]{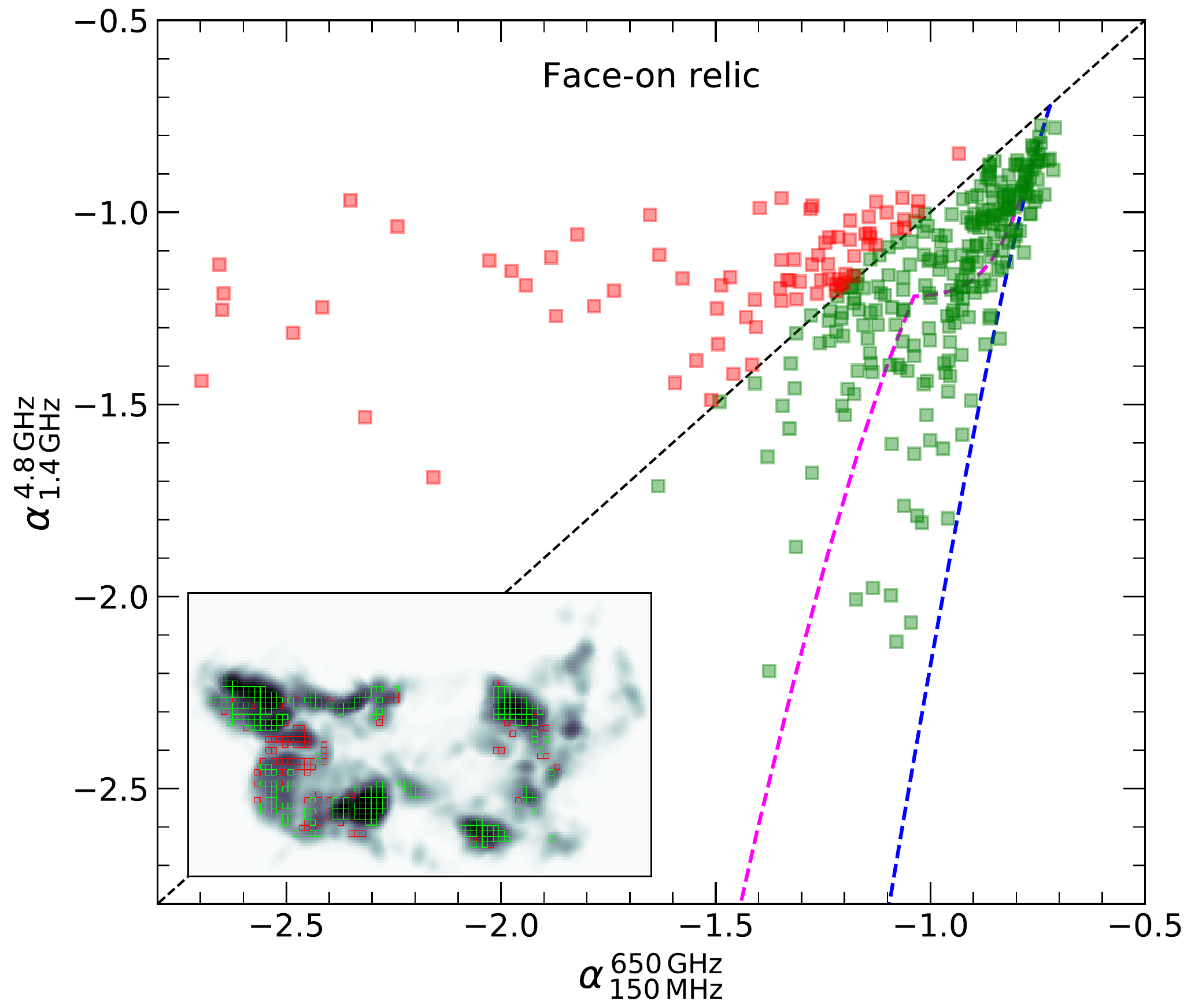}
 \includegraphics[width = 0.45 \textwidth]{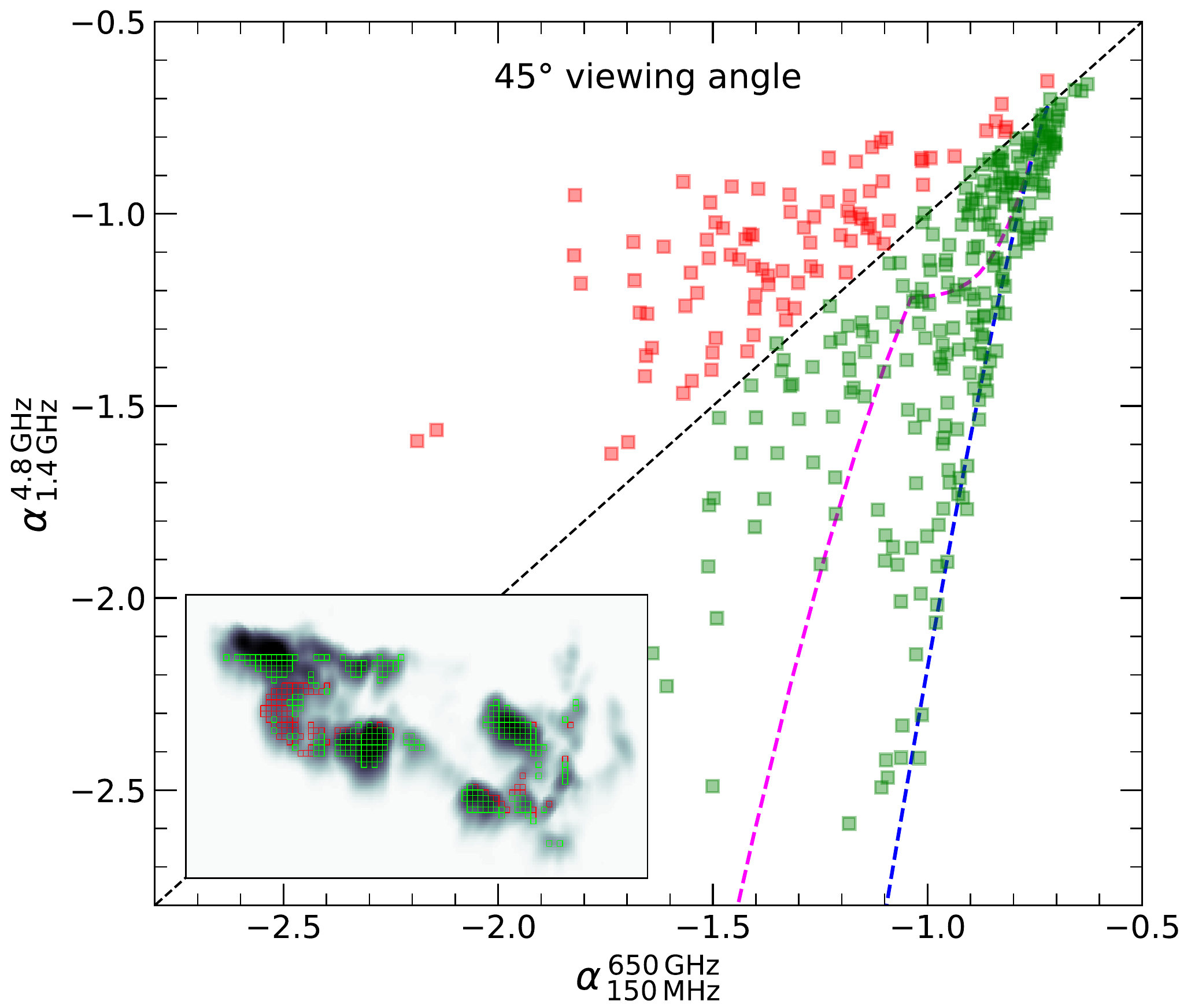}
\includegraphics[width = 0.45 \textwidth]{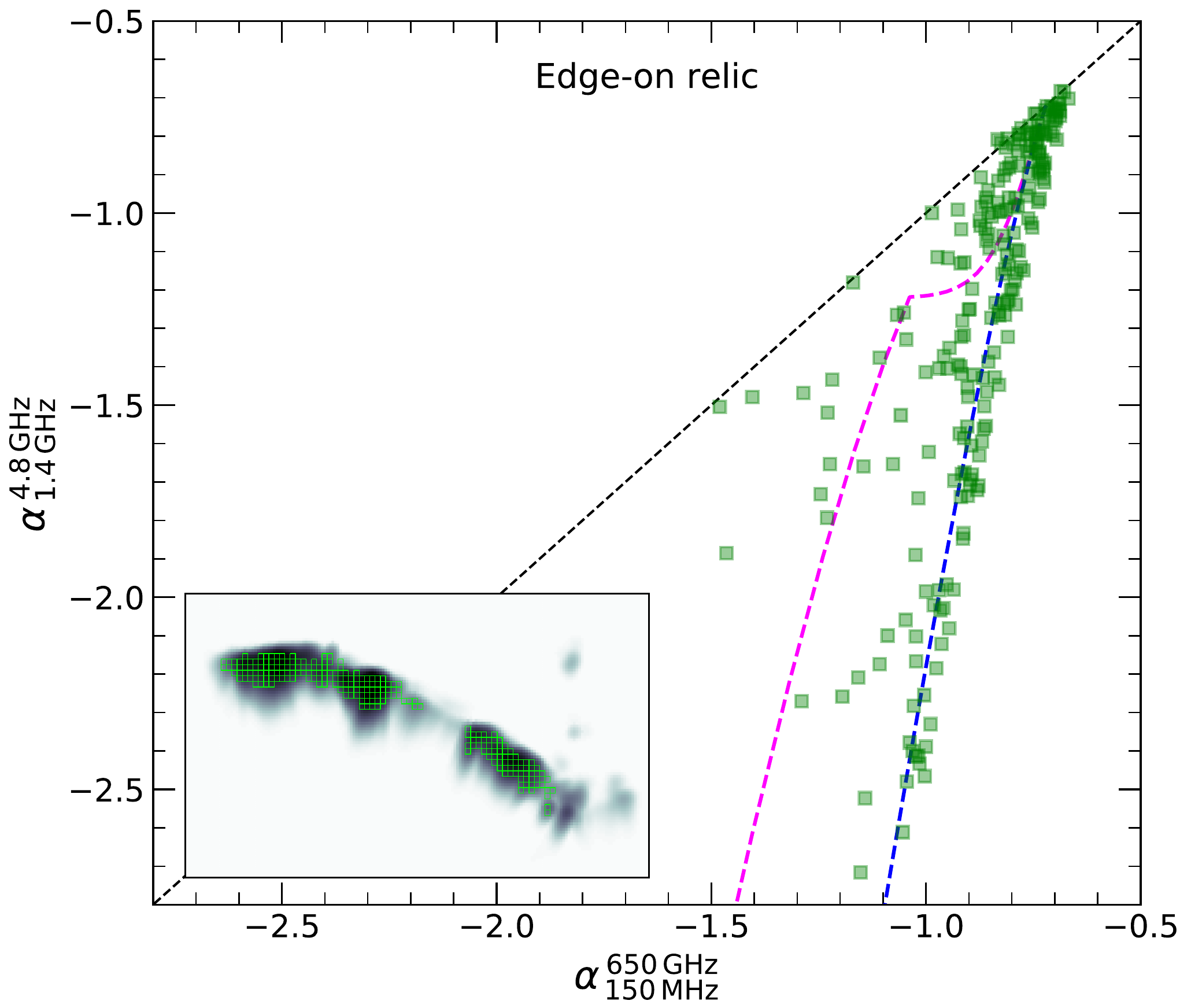}
\vspace{-0.2cm}
\caption{Radio color-color plots of the simulated relic from \citet{2019arXiv190911329W} at different viewing angles, superimposed with the JP (blue) and KGJP (magenta) spectral aging models obtained assuming $\alpha_{\rm inj}= -0.70$. The top panel shows the face-on view of the relic, the middle panel at $45\degree$ viewing angle, and the bottom panel the edge-on view. The spectral indices are measured inside the $\sim (15.8 \ \mathrm{kpc})^2$ boxes, as shown in the insets. The colors of the boxes match the colors of the points in the color-color plot; a red square is above the power-law line and a green square is below the power-law line. The plots reflect the differences in the spectral shapes of relics depending on the viewing angle.}
\label{fig::simulation_colorcolor}
\end{figure}

Adopting the standard DSA model, we can constrain the acceleration efficiency for the relic. The synchrotron power of the relic is supplied by a small fraction ($\xi_e$) of the kinetic energy flux ($\Phi_k$) of the underlying shock, which is assumed to inject relativistic electrons \citep{Hoeft2007}. For simplicity, we model the relic with surface area $L^2$, where $L$ is set to the largest linear scale such that $L=0.85$~Mpc. The accelerated seed electrons are supposed to be come from the thermal pool of the ICM at the relic location, which we characterize with a temperature $T_e=12$~keV and an electron density $n_e=10^{-3}\rm cm^{-3}$ \citep{vanWeeren2017b}.

The results are shown in Fig.~\ref{figxyz}. By excluding unrealistically large values of the magnetic field ($\geq 10 ~\rm \mu G$), we can thus conclude that $\xi_e$ needs to be larger than $\simeq 3\times 10^{-3}$ . For a few $\upmu$G magnetic field, the DSA requires values of $\xi_e \sim10^{-3} - 10^{-1}$. This range of value is well in line with what is required for radio relics observed at $\sim \rm GHz$ frequency \citep{2015MNRAS.451.2198V, 2020A&A...634A..64B}. 

However, theoretical and observational studies indicate that in the framework of DSA, cosmological shocks can reach $\xi_e$ up to a few oder of $\sim10^{-3}$ \citep[e.g.,][]{2016MNRAS.459...70V,2017MNRAS.470..240N,2020A&A...634A..64B}. It has thus been suggested that the high acceleration efficiency required to match observed radio powers requires previous acceleration events, for example from the crossing of previous structure formation shocks, or by the injection of fossil electrons by radio galaxies  \citep[re-acceleration models, see][]{Markevitch2005,Pinzke2013,Kang2016a,2015MNRAS.451.2198V, 2020A&A...634A..64B}. Note that the magnetic field is degenerate with the (unknown) electron acceleration efficiency $\xi_e$. However, we can set a lower limit to $\xi_e$. If the shock strength corresponds to the radio spectrum according to Equation~\ref{JEeq::alpha_int_mach} (from the standard DSA), an acceleration efficiency below 10\,\%could reach the observed luminosity values only if the magnetic field strength is $B<5\,\upmu \rm G$ and a large fraction of the shock surface shows the derived Mach number ($3.5-4$) or higher. 

To provide an alternative magnetic field estimate, we can make the assumption that the sum of the energy density in magnetic field $U_B$ and relativistic particles $U_{CR}$ is minimized \citep{1980ARA&A..18..165M}. Classical methods to compute $U_{CR}$ that deal with the injection spectrum of the shocked particles do not account for the energy losses of the electrons in the plasma \citep{1997A&A...325..898B, 2005AN....326..414B}. Instead, we include them in the computation by linking the CR energy to the shock kinetic energy through the electron acceleration efficiency $\xi_{e}$ (in a similar manner to the DSA modeling above) and the ratio of energy densities between protons and electrons $k$; see \citet{Locatelli2020} for details. We thus write:

\begin{equation}
1/2\, \rho_u\,v^3_u/v_d \, \xi_e\, (1+k) = B^2 /(8\pi),
\end{equation}
where the gas density $\rho$ and shock velocity $v$ have been computed either upstream ($u$) or downstream ($d$) of the shock front using Mach number $\mathcal{M}=3.7$. Values of $\xi_e$ that minimize the total energy density $U_B+U_{CR}$ lie within the range $[9\times 10^{-4};\,2\times 10^{-2}]$ for $k\in [0;\,1000]$, where $k=0 (1000)$ stands for a gas with energy density dominated by electrons (protons).

Diffusive particle acceleration typically predicts $k$ of the order of 100 \citep{Bell1978}. In this way, we retrieve $B=2.5-9\,\upmu\rm G$ for $0<k<100$ by requiring $\xi_e\,(1+k)<<1$. (see the equipartition dotted lines in Fig.~\ref{figxyz}). The values of magnetic field derived in this way are consistent with the ones from the DSA modeling. We also show the expected X-ray flux in the $20-80$~keV band by inverse Compton scattering (IC) given the electron population and magnetic field responsible of the radio emission. Magnetic fields as low as $B\simeq 1\,\upmu$G would imply IC radiation compatible with the current upper limits.

\section{Comparison with simulations}
\label{simlation_part}

We compare MACS\,J0717.5+3745 with the radio emission of a simulated galaxy cluster already studied by \citet{2019arXiv190911329W} and discussed in \citet{Rajpurohit2020a}. Since the simulation allows us to chose arbitrary line-of-sight, in this work we examined different viewing angles, and found a radio morphology that is qualitatively similar to MACS\,J0717.5+3745. We note that the simulation is not meant to exactly reproduce MACS\,J0717.5+3745.

This simulation was performed with the cosmological magneto-hydrodynamical ENZO code \citep[][]{ENZO_2014} and it belongs to a recent study of magnetic field dynamo amplification in high-resolution simulations of galaxy clusters \citep[][]{2019MNRAS.486..623D}. The synthetic radio model given by this simulation is the most detailed to date, as it includes complex magnetic field patterns formed during the cluster merger.  It also models the Mach number dependence of radio-emitting electrons following DSA. The model also accounts for the downstream cooling of accelerated electrons in the downstream region of the shock front. 

For numerical details, we refer readers to \citet{2019arXiv190911329W}. The radio power of the simulated relic, namely  $\sim 1.32 \times 10^{30} \ \mathrm{erg/s/Hz}$ at $1.4 \ \mathrm{GHz}$, is about a factor of $\sim 800$ dimmer than MACS\,J0717.5+3745.  Since we are particularly interested in the radio spectrum of the observed relic,  rather than the absolute power, the simulated relic remains suitable for such a comparison.

In order to investigate the role of projection and viewing angle, we studied the relic seen face-on and at a $45\degree$ viewing angle. We calculate the integrated spectrum of the whole simulated relic and three non-overlapping subregions as shown in Fig.\,\ref{fig::simu_spectrum}. As in real observations, the simulated radio emission shows a broadband power law spectrum with index of 
$\alpha \sim -1.17$, independently of the viewing angle.  In addition, the subregions show similar spectral slopes. Also seen edge-on, the simulated relic shows a similar spectral slope of $\sim -1.16$ across the whole relic and in three individual subregions \citep[][]{Rajpurohit2020a}. This suggests that radio power from each of the simulated subregions is dominated by shocks with Mach numbers ${\mathcal{M} \sim 3.4-3.7}$. Once again, we remind the readers that these numbers lie in the tail of the whole distribution of Mach numbers available at the shock front.

To understand the curvature distribution of the relic in MACS\,J0717.5+3745, we performed a color-color analysis on the simulated relic when seen at different viewing angles. The corresponding plot is given in Fig. \ref{fig::simulation_colorcolor}. We compute the low, $144-650$\,MHz, and high, $1.4-4.85$ \,GHz, frequency spectral indices in small boxes with a physical size of $\sim (15.8 \ \mathrm{kpc})^2$ as shown in the inset of Fig. \ref{fig::simulation_colorcolor}. The points that are below the line are measured inside the green boxes. The red points indicate the regions where the high-frequency spectral index is larger than the low-frequency index.

There is a striking difference between integrated spectra (obtained by averaging over big regions, Fig.\,\ref{fig::simu_spectrum} ) and spectra visible in the color-color plots (derived from much smaller regions, Fig. \ref{fig::simulation_colorcolor}). The former is a simple power-law while the latter shows complex curvature and possibly more than one component. The regions used for  extracting the broadband spectra are large enough to encompass the full cooling region behind the shock. As a result the combination of all the different cooling spectra broadens the integrated spectrum to a power law. In contrast, the small boxes used for creating the color-color plots are  too small to catch the full cooling region behind each shock, so we see a wide range of curvatures depending on the particular mixture along each line of sight. 

For a face-on relic, we find a broad range of spectra. Some of the data points are clustered around the black dashed-line that gives equality between the low and the high frequency spectral indices; see the top panel of Fig.\,\ref{fig::simulation_colorcolor}. This indicate power-law spectra with no or very little curvature. Majority of points (shown with green) fall below the black line, indicating a convex curvature. On the other hand, the red points that are significantly above the black line, indicate concave spectra. As shown in the middle panel of Fig.\,\ref{fig::simulation_colorcolor}, the relic seen at $45\degree$ also show points that lie above the power-law line. Only the projection of different emitting structures can explain such a behavior. From the map, we deduce that the regions with such concave spectra fall into the transition regions where the different radio structures connect and overlap. In this case, the relic also shows a significant negative curvature. 
 
The trajectory of the color-color plot of the same relic when seen edge-on is shown in the bottom panel of Fig.\,\ref{fig::simulation_colorcolor}. In the edge-on view, a clear trend of negative curvature is visible. The trajectory in the simulated color-color plot resembles those reported for the Toothbrush and for the Sausage relic \citep{vanWeeren2012a,Stroe2013,Rajpurohit2020a}, which are most likely seen edge-on . These differences in the color-color plots highlight the possible differences in the spectral shapes of relics depending on the viewing angle. 

We also overplot the JP and KGJP models to the simulated color-color plots. The JP model can be used when a (planar) shock front is parallel to the line of sight and the KGJP when the (finite and planar) shock front is somewhat inclined to the line of sight, see \cite{Rajpurohit2020a} for discussion. For an edge-on relic, the JP model provides a very good match; see the bottom panel of Fig.\,\ref{fig::simulation_colorcolor}. This indicates that every line of sight mostly probes a specific electron population, characterized by a single age. At $45\degree$ viewing angle also, some points in the color-color plot along the same locus seems to follow the JP model.  

We emphasize here that in the edge-on view, the scatter around the JP line is small.  It may imply that there is a similar shape to the relativistic electron distribution throughout the relic, with different spectral indices observed when the magnetic field varies or different amounts of radiative and adiabatic losses have taken place. This suggest that conditions along each line of sight are relatively homogeneous in this view. By contrast, the large scatter in curvatures seen in the face-on and $45\degree$ viewing angle is due to the fact that along any of these lines of sight there is a substantial mixture of both magnetic field strengths and loss histories, resulting in broader and even concave total spectra.

The curvature trend observed for the relic in MACS\,J0717.5+3745 (see the left panel of Fig.\,\ref{CC_plot}) is different from these three cases. Excluding the overlapping region between the NAT and R3, all points lie below the power-law line. The observational color-color plot does not show any hint of positive curvature. Thus, the viewing angle of the relic is mostly likely above $45\degree$. However, none of the spectral aging models fits the observed data in MACS\,J0717.5+3745. We thus speculate that the relic is likely to be inclined along the line of sight, and that projection effects might be responsible for the observed spectral shape in MACS\,J0717.5+3745.


\section{Summary and Conclusions}
\label{summary}

In this work, we have presented new uGMRT (300-850\,MHz) radio observations of the galaxy cluster  MACS\,J0717.5+3745. These observations were combined with the new LOFAR and VLA observations in order to carry out a detailed spectral analysis.  We summarize the overall results as follows:

\begin{enumerate} 

\item{} Our new uGMRT  images  confirm the presence of  sub-structures across the relic and the halo, previously reported at the high-frequencies. The new LOFAR observation reveals that the tails of the NAT, located at the cluster center, bend to the south rather than fading into the relic. \\

\item{} The spatially integrated radio emission from the relic, as well as from each of its subregions, is a simple power law from 144\,MHz to 5.5\,GHz.  The power law spectral indices between  $-1.13$  to $-1.18$, indicates that the underlying shocks have Mach numbers ${\mathcal{M} \sim3.5 - 4.0}$. As found for the Toothbrush relic \citep{Rajpurohit2020a}, this suggests that the high-Mach number tail of the shock distribution governs the integrated radio spectral index of relics.\\

\item{}The close comparison to recent numerical simulations of radio relics suggest that spectra which are integrated over sufficiently large regions of the relic show a similar slope, independent of viewing angle. Both observations and simulation suggest that relics contain both weak and strong shocks, but that shocks with Mach number ${\mathcal{M}\simeq 3.5 - 4.0}$ dominate the relic radio spectrum.\\

\item{} The injection index measured from the spectral index maps between 144\,MHz and 1.5\,GHz, varies between $-0.70$ and $-0.90$, suggesting inhomogeneities in the Mach number. The spectral shapes inferred from spatially resolved regions reveal spectral curvature. However, none of the spectral aging models fit the observed data. We speculate that the relic is inclined along the line-of-sight. \\

\item{} The color-color plots obtained with simulations show different spectral shapes. The locus of points in the simulated color-color plots changes significantly with the relic viewing angle. This shows that the relativistic electron distribution has a common shape throughout the relic when conditions are relatively homogeneous along each line of sight, as in the edge on view.\\

\item{} We find that there are structures in the relic with different spectral indices overlapping with each other, indicating that the relic is composed of multiple fine filaments.\\ 

\item{} If CRe are accelerated from the thermal ICM, a plausible acceleration efficiency below 10\% results in the observed radio power of the relic if the magnetic field is below $5\upmu \rm G$ and a large fraction of the shock shows high Mach number. \\

\item{} Based on the new observations, we raise the possibility that the NAT and R3 could be two different structures projected along the same line of sight. In this scenario the relic is not powered by the shock re-acceleration of fossil electrons by the NAT.\\

\end{enumerate}

\begin{acknowledgements}
KR, FV, NL, and PDF acknowledges financial support from the ERC Starting Grant "MAGCOW", no. 714196. DW is funded by the Deutsche Forschungsgemeinschaft (DFG, German Research Foundation)-441694982. RJvW and AB acknowledges support from the VIDI research programme with project number 639.042.729, which is financed by the Netherlands Organization for Scientific Research (NWO). WF acknowledges support from the Smithsonian Institution and the High Resolution Camera Project through NASA contract NAS8-03060. AB, MB, CJR, and EB acknowledges financial support from the ERC Starting Grant ``DRANOEL" number 714245. GB and FG acknowledge support from INAF mainstream program ``Galaxy clusters science with LOFAR". AD acknowledges support by the BMBF Verbundforschung under grant 05A17STA.  Partial support for LR comes from U.S. National Science Foundation grant AST17-14205 to the University of Minnesota. Basic research in radio astronomy at the Naval Research Laboratory is supported by 6.1 Base funding.
A part of the data reduction was performed using computer facilities at Th\"uringer Landessternwarte Tautenburg, Germany. The cosmological simulations in this work were performed using the ENZO code (http://enzo-project.org). The authors gratefully acknowledge the Gauss Centre for Supercomputing e.V. (www.gauss-centre.eu) for supporting this project by providing computing time through the John von Neumann Institute for Computing (NIC) on the GCS Supercomputer JUWELS at J\"ulich Supercomputing Centre (JSC), under projects no. HHH42 and {\it stressicm} (PI F.Vazza) as well as HHH44 (PI D. Wittor). We also acknowledge the usage of online storage tools kindly provided by the INAF Astronomical Archive (IA2) initiave (http://www.ia2.inaf.it). This research made use of computer facility  on the HPC resources at the Physical Research Laboratory (PRL), India.

We thank the staff of the GMRT that made these observations possible. GMRT is run by the National Centre for Radio Astrophysics of the Tata Institute of Fundamental Research.
LOFAR \citep{Haarlem2013} is the Low Frequency Array designed and constructed by ASTRON. It has observing, data processing, and data storage facilities in several countries, which are owned by various parties (each with their own funding sources), and that are collectively operated by the ILT foundation under a joint scientific policy. The ILT resources have benefited from the following recent major funding sources: CNRS-INSU, Observatoire de Paris and Universit\'{e} d'Orl\'{e}ans, France; BMBF, MIWF-NRW, MPG, Germany; Science Foundation Ireland (SFI), Department of Business, Enterprise and Innovation (DBEI), Ireland; NWO, The Netherlands; The Science and Technology Facilities Council, UK; Ministry of Science and Higher Education, Poland; The Istituto Nazionale di Astrofisica (INAF), Italy. This research made use of the LOFAR-UK computing facility located at the University of Hertfordshire and supported by STFC [ST/P000096/1], and of the LOFAR-IT computing infrastructure supported and operated by INAF, and by the Physics Dept. of Turin University (under the agreement with Consorzio Interuniversitario per la Fisica Spaziale) at the C3S Supercomputing Centre, Italy.

The National Radio Astronomy Observatory is a facility of the National Science Foundation operated under cooperative agreement by Associated Universities. 
\end{acknowledgements}

\bibliographystyle{aa}

\bibliography{MACSJ0717a.bib}

\end{document}